\documentclass[10pt]{article}
\usepackage{amsmath}
\usepackage{amssymb}

\newcommand{\bbbone}{\mathchoice {\rm 1\mskip-4mu l} {\rm 1\mskip-4mu l}
{\rm 1\mskip-4.5mu l} {\rm 1\mskip-5mu l}}
\newcommand{\scalprod}[2]{\left\langle {#1}, {#2}\right\rangle}
\newcommand{\dom}{{\cal D}}
\newcommand{\RE}{{\rm Re}}
\newcommand{\IM}{{\rm Im}}
\newcommand{\fer}[1]{(\ref{#1})}
\newcommand{\ran}{{\rm Ran\,}}
\renewcommand{\ker}{{\rm Ker\,}}

\newcommand{\h}{{\cal H}}
\newcommand{\cx}{{\mathbb C}}
\newcommand{\rx}{{\mathbb R}}

\renewcommand{\r}{{\rx}}

\newcommand{\supp}{{\rm supp}}
\newcommand{\Lbar}{{\overline L}}

\newcommand{\bk}{{\boldsymbol k}}
\newcommand{\bl}{{\boldsymbol l}}
\newcommand{\ux}{{\underline x}}
\newcommand{\uk}{{\underline k}}

\renewcommand{\d}{{\rm d}}
\newcommand{\Lbarlambda}{\overline L_\lambda}
\newcommand{\Kbarlambda}{\overline K_\lambda}

\newcommand{\sgn}{{\rm sgn}}

\newcommand{\hh}{{\frak H}}

\newcommand{\mm}{{\frak M}}

\newcommand{\cc}{{\cal C}}
\newcommand{\ff}{{\cal F}}
\newcommand{\Pbar}{{\, \overline{\!P}}}

\newcommand{\tr}{{\rm tr\,}}

\renewcommand{\k}{{\cal K}}

\newcommand{\hr}{{\cal H}_{\rm R}}

\renewcommand{\b}{{\cal B}}
\newcommand{\e}{{\rm e}}
\newcommand{\Fplus}{{\cal F}_{\!\!+}}

\renewcommand{\i}{{\rm i}}

\newcounter{resultcounter}[section]

\newtheorem{theorem}[resultcounter]{Theorem}
\newtheorem{lemma}[resultcounter]{Lemma}
\newtheorem{proposition}[resultcounter]{Proposition}

\begin{document}

\setcounter{page}{0}

\title{Level shift operators for open quantum systems}

\author{
Marco Merkli \footnote{Supported by a CRM-ISM postdoctoral fellowship and by McGill University; merkli@math.mcgill.ca; http://www.math.mcgill.ca/$\sim$merkli/}\ \\
Dept. of Mathematics and Statistics, McGill University\\
805 Sherbrooke  W., Montreal\\
Canada, QC, H3A 2K6\\
\vspace{-.3cm}
\centerline{and}
\and
Centre de Recherches Math\'ematiques,
Universit\'e de Montr\'eal\\
Succursale centre-ville, Montr\'eal\\
Canada, QC, H3C 3J7
}
\date{\today}
\maketitle
\vspace{-1cm}

\begin{abstract}
Level shift operators describe the second order displacement of
eigenvalues under perturbation. They play a central role in
resonance theory and ergodic theory of open quantum systems at
positive temperatures.

We exhibit intrinsic properties of level shift operators,
properties which stem from the structure of open quantum
systems at positive temperatures and which are common to all such
systems. They determine the geometry of resonances bifurcating
from eigenvalues of positive temperature Hamiltonians and they
relate the Gibbs state, the kernel of level
shift operators, and zero energy resonances.

We show that degeneracy of energy levels of the small part of the open quantum system causes the Fermi Golden Rule Condition to be violated and we
analyze ergodic properties of such systems.

\end{abstract}

\setcounter{page}{1}
\setcounter{section}{1}

\setcounter{section}{0}

\section{Introduction and main results}

Level shift operators emerge naturally in the context of
perturbation theory of (embedded) eigenvalues, where they govern
the {\it shifts of levels} (resonances) at second order in
perturbation. They play a central role in many recent works on ergodic properties of open quantum systems at positive
temperature \cite{JP1,JP2,BFS,DJ,M2,FM1,FM2,FMS,JP3,MMS,AF,A}. The
dynamics of such systems is an automorphism group of the algebra
of observables generated by an operator $L_\lambda =L_0+\lambda
I$, where the selfadjoint $L_0$ describes the free
dynamics of two (or more) uncoupled subsystems, $\lambda\in \r$ is
a coupling constant, and $I$ is an interaction operator.

Ergodic properties are encoded in the spectrum of $L_\lambda$. If the system has an equilibrium state $\Omega_{\beta,\lambda}$ at positive temperature $1/\beta$ one
shows that $L_\lambda\Omega_{\beta,\lambda}=0$ and that if $\ker
L_\lambda =\cx\Omega_{\beta,\lambda}$ then any state initially
close to equilibrium approaches the equilibrium state in the limit of large times. (We do not
address the question of mode or speed of the return to equilibrium in this
outline).

It follows from the algebraic structure of quantum systems at positive temperatures that the
operator $L_0$ has necessarily a degenerate kernel whose elements are in
one-to-one correspondence with invariant states of the uncoupled
system. In order to prove return to equilibrium one needs to show
that the degeneracy of the eigenvalue zero, which is embedded in
continuous spectrum, is lifted under perturbation: $\dim \ker
L_\lambda =1$ for $\lambda\neq 0$. This has been proven for
several concrete models \cite{JP1,JP2,BFS,M2,DJ,FM2,AF}. The fate
of embedded eigenvalues under perturbation can be described by
spectral resonance theory if the system has certain deformation
analyticity properties, \cite{JP1,JP2,BFS,AF}, or, for less
regular systems, by a Mourre theory \cite{DJ} or by a positive
commutator theory, \cite{M2,FM2}.

A core strategy common to these methods is to reduce the spectral
analysis of the operator $L_\lambda$ around the origin to that of
a reduced operator acting on a smaller Hilbert space (which is
finite-dimensional in all works cited above). This procedure
is sometimes called `integrating out degrees of freedom'. For
deformation analytic systems it can be implemented by applying the
so-called Feshbach map $F$ \cite{BFS1} to a suitably deformed
operator $L_\lambda(\tau)$, where $\tau\in\cx$ is the deformation
parameter \footnote{ The operator $L_\lambda$ itself is typically
not in the domain of the map $F$ but $L_\lambda(\tau)$ is for
$\tau\not\in\mathbb R$. Arguments similar to the ones we give here
 also work for systems which are not deformation analytic, but they
need a technically more elaborate presentation.} . The Feshbach
map has an isospectrality property, implying that the kernels of
$F(L_\lambda(\tau))$ and of $L_\lambda$ are isomorphic. An
expansion in the coupling constant $\lambda$ gives
\begin{equation}
F(L_\lambda(\tau))= -\lambda^2\Lambda_0+O(\lambda^3),
\label{*}
\end{equation}
where the operator $\Lambda_0$ is independent of the deformation
parameter $\tau$. The property of return to equilibrium follows
 if $\Lambda_0$ has simple kernel
because then \fer{*} and the isospectrality of the Feshbach map imply that for small $\lambda\neq 0$,
$\dim\ker L_\lambda\leq 1$, and since
$L_\lambda\Omega_{\beta,\lambda}=0$ one must have $\ker
L_\lambda=\cx \Omega_{\beta,\lambda}$. The operator $\Lambda_0$
is called the level shift operator (associated to the eigenvalue
zero of $L_\lambda$). For a more detailed description of the
emergence of level shift operators in perturbation theory we refer
to the works cited above, and also to \cite{DF}.

Level shift operators are equally important in the study of
systems far from equilibrium, where the system does not have an
equilibrium state, for instance when several thermal reservoirs at
different temperatures are coupled, \cite{JP3,MMS,A}, or when the
small system does not admit an equilibrium state, \cite{FM1,FMS}.
In what follows we discuss the former case. The role of the
equilibrium state is now played by a reference state
$\psi_\lambda$, e.g. the product state of the small system and the
reservoirs in equilibria at different temperatures. An interaction
operator $W$ can be chosen such that the Heisenberg dynamics of
the system is generated by the operator $K_\lambda =L_0+\lambda
W$, satisfying $K_\lambda\psi_\lambda=0$. Unlike $I$ in the
situation of systems close to equilibrium, the operator $W$ cannot
be chosen to be normal. A detailed spectral analysis of operators
of this type (called `C-Liouville operators') is carried out in
\cite{MMS}. One obtains the level shift operator $\Lambda_0$
associated to the eigenvalue zero of $K_\lambda$ by an expansion
of $F(K_\lambda(\tau))$ in $\lambda$, as in the situation above.
In the context of systems far from equilibrium, a dynamical
resonance theory shows that if $\Lambda_0$ has simple kernel then
the system possesses a unique time-asymptotic limit state (for
small $\lambda$), which is a non-equilibrium stationary state.

We should also point out that level shift operators have a
dynamical interpretation as ``Davies generators'' of the reduced
dynamics in the van Hove limit, \cite{DJ,DF1}.

\medskip

In the present paper we examine properties of level shift
operators which do not depend on particularities of the system in
question, but which originate from the {\it structure} common to
all open quantum systems at positive temperatures. One of the main
ingredients determining this structure is the Tomita--Takesaki
theory of von Neumann algebras.

We describe the geometry of resonances bifurcating from real
eigenvalues of Liouville operators in Theorem \ref{thm3}. Part (c) of that theorem examines the role of the degeneracy
of energy levels of the small system. The interplay between the
Gibbs state of the small system and the kernel of the level shift
operator is described in Theorem \ref{thm2}. Some of these
{\it intrinsic} properties we exhibit here have been observed in the analysis of
specific systems carried out in the references given above.

Apart from an analysis of the structure of level shift operators
we study in Section \ref{degspec} the {\it dynamics} of systems where a
Bosonic heat reservoir is coupled to a small system whose Hamiltonian
has degenerate eigenvalues. For such systems the so-called Fermi
Golden Rule Condition is violated due to the fact that the
interaction ``cannot couple the Bohr frequency zero'' (arising
from the degenerate levels) to the zero energy reservoir modes in
an effective way. We prove return to equilibrium for such systems
by taking into account higher order corrections in the
perturbative spectral analysis of the operator $L_\lambda$. A
consequence of the degeneracy of energy levels of the small system
is that the approach to equilibrium is still exponentially fast
but with relaxation time of $O(\lambda^{-4})$ as opposed to the
shorter relaxation time $O(\lambda^{-2})$ for systems with
non-degenerate spectrum.

\medskip

The organization of this paper is as follows. In Section
\ref{subs1} we summarize some facts about open quantum systems.
Our main results concerning the structure of level shift operators
are Theorems \ref{thm3}, \ref{thm2} and \ref{thm5}, given in
Section \ref{subsmainresult}. In Section \ref{exsect} we present
examples of concrete models to which our results apply. We analyze
systems with Hamiltonians having degenerate eigenvalues in Section
\ref{degspec}. Section \ref{proofsection} contains proofs.

\subsection{Open quantum systems}
\label{subs1}

A detailed description of open quantum systems can be found in \cite{BR}, in the above-mentioned works, and also in \cite{JP,M1}.

Consider a quantum system possessing finitely many degrees of freedom, like a single particle or a molecule, or a system with finitely many energy levels. We denote by $\k$ the Hilbert space of pure states of this ``small system'' and by $H$ its Hamiltonian. We allow the case $\dim \k=\infty$ but require that the {\it Gibbs state} $\Psi_\beta$ exists, i.e., that $\e^{-\beta H}$ is trace class on $\k$, for some inverse temperature $0<\beta<\infty$. In the above example the particle or molecule must thus be confined, e.g. by a potential.

We view the Gibbs state $\Psi_\beta$ as a vector in the Hilbert space $\h_{\rm S}=\k\otimes\k$. The algebra of all bounded operators, $\b(\k)$, contains the observables of the small system. It is represented on $\h_{\rm S}$ as the von Neumann algebra $\mm_{\rm S}=\b(\k)\otimes\bbbone_\k\subset \b(\h_{\rm S})$, and the dynamics on $\mm_{\rm S}$ is implemented by $t\mapsto \e^{\i t L_{\rm S}}A\e^{-\i tL_{\rm S}}$, where
\begin{equation}
L_{\rm S}= H\otimes\bbbone_\k - \bbbone_\k\otimes H
\label{a13}
\end{equation}
is called the standard Liouville operator of the small system. The Gibbs vector $\Psi_\beta$ is cyclic and separating for $\mm_{\rm S}$. We denote the modular conjugation associated to the pair $(\mm_{\rm S},\Psi_\beta)$ by $J_{\rm S}$. The standard Liouville operator $L_{\rm S}$ satisfies the relations $L_{\rm S}\Psi_\beta=0$ and $J_{\rm S}L_{\rm S}J_{\rm S}=-L_{\rm S}$.

In models of systems close to equilibrium, the small system is placed in a (single) environment (reservoir) modeled by a ``large'' quantum system having infinitely many degrees of freedom. A common example is a spatially infinitely extended ideal quantum gas (of Bosons or Fermions). We assume that the reservoir has an equilibrium state (for some inverse temperature $0<\beta<\infty$) which is represented by the vector $\Phi_\beta$ in the reservoir Hilbert space $\h_{\rm R}$.

Observables of the reservoir are operators belonging to (or affiliated with) a von Neumann algebra $\mm_{\rm R}\subset\b(\h_{\rm R})$. Their dynamics is given by a group of automorphisms of $\mm_{\rm R}$, $t\mapsto \e^{\i tL_{\rm R}} A \e^{-\i tL_{\rm R}}$, generated by a selfadjoint standard Liouville operator $L_{\rm R}$. Being a KMS vector w.r.t. this dynamics, $\Phi_\beta$ is cyclic and separating for $\mm_{\rm R}$. We denote the modular conjugation associated to $(\mm_{\rm R},\Phi_\beta)$ by $J_{\rm R}$. The operator $L_{\rm R}$ has the properties $J_{\rm R}L_{\rm R}J_{\rm R}=-L_{\rm R}$ and $L_{\rm R}\Phi_\beta=0$.

\medskip
The von Neumann algebra $\mm=\mm_{\rm S}\otimes\mm_{\rm R}$, acting on the Hilbert space $\h=\h_{\rm S}\otimes\h_{\rm R}$, contains the observables of the combined system. Elements in this algebra evolve according to the group of automorphisms of $\mm$ generated by the selfadjoint operator
\begin{equation}
L_0=L_{\rm S}+L_{\rm R}.
\label{a.3.8}
\end{equation}
The vector
\begin{equation}
\Omega_{\beta,0}=\Psi_\beta\otimes\Phi_\beta
\label{a3.9}
\end{equation}
defines the equilibrium state w.r.t. this dynamics, at inverse temperature $\beta$.

To describe the interacting dynamics we introduce the map $\pi$ which sends linear operators on $\k\otimes\hr$ to linear operators on $\k\otimes\k\otimes\hr$ according to
\begin{equation}
\pi(A\otimes B) = A\otimes\bbbone_\k\otimes B.
\label{a6}
\end{equation}
The interaction between the small system and the reservoir is specified by a selfadjoint operator $v$ on $\k\otimes\h_{\rm R}$ such that
\begin{equation}
V=\pi(v)
\label{a.3.91}
\end{equation}
is a selfadjoint operator affiliated with $\mm$. Let $J=J_{\rm S}\otimes J_{\rm R}$. We assume that
\begin{itemize}
\item[{\bf B1}] $L_0+\lambda V$ is essentially selfadjoint on $\dom(L_0)\cap\dom(V)$ and $L_0+\lambda V-\lambda JVJ$ is essentially selfadjoint on $\dom(L_0)\cap\dom(V)\cap\dom(JVJ)$, where $\lambda$ is a real coupling constant.
\end{itemize}
If assumption B1 holds then the selfadjoint operator
\begin{equation}
L_\lambda = L_0+\lambda I,
\label{a4}
\end{equation}
where
\begin{equation}
I = V - J V J,
\label{a5}
\end{equation}
generates a group of automorphisms $\alpha^t_\lambda = \e^{\i tL_\lambda}\cdot \e^{-\i tL_\lambda}$ of $\mm$ (see e.g. \cite{DJP}, Theorem 3.5). We are interested in interactions for which the coupled dynamics $\alpha_\lambda^t$ admits an equilibrium state. It is known that the condition
\begin{itemize}
\item[{\bf B2}] $\Omega_{\beta,0}\in \dom(\e^{-\beta (L_0+ \lambda V)/2})$
\end{itemize}
implies that
\begin{equation}
\Omega_{\beta,\lambda}=\frac{\e^{-\beta(L_0+\lambda V)/2}\Omega_{\beta,0}}{\|\e^{-\beta(L_0+\lambda V)/2}\Omega_{\beta,0}\|}
\label{a20}
\end{equation}
is a $(\beta,\alpha_\lambda^t)$--KMS state, and that the following properties hold: $JL_\lambda J=-L_\lambda$, $L_\lambda\Omega_{\beta,\lambda}=0$, and  $\lim_{\lambda \rightarrow 0}\|\Omega_{\beta,\lambda}-\Omega_{\beta,0}\|=0$ (see e.g. \cite{DJP}, Theorem 5.5).

\medskip

In models for systems far from equilibrium the small system is coupled to several reservoirs. The Hilbert space of the small system plus $R\geq 2$ reservoirs is ${\cal H}={\cal H}_{\rm S}\otimes {\cal H}_{\rm R}\otimes\cdots\otimes {\cal H}_{\rm R}$ and the non-interacting dynamics of the algebra $\mm=\mm_{\rm S}\otimes\mm_{\rm R}\otimes\cdots\mm_{\rm R}$ is generated by the selfadjoint standard free Liouville operator
\begin{equation}
L_0= L_{\rm S} + \sum_{r=1}^R L_{{\rm R},r},
\end{equation}
where $L_{{\rm R},r}$ acts non-trivially only on the $r$-th reservoir space.
The interaction is determined by an operator
\begin{equation}
V=\sum_{r=1}^R\pi(v_r),
\label{n1}
\end{equation}
where $v_r$ is the selfadjoint operator on ${\cal K}\otimes{\cal H}_{\rm R}$ representing the coupling between the small system and reservoir $r$. We understand that $\pi(v_r)$ acts trivially on all reservoir Hilbert spaces except the $r$-th one. As mentioned in the introduction, the role of the equilibrium state is now played by a reference state $\psi_\lambda\in{\cal H}$. We assume the following.
\begin{itemize}
\item[{\bf C}] The interacting dynamics of $\mm$ is generated by the operator $K_\lambda = L_0+\lambda W$, where $W$ is a bounded, linear (generally non-symmetric) operator on ${\cal H}$, s.t. $K_\lambda\psi_\lambda=0$ for all $\lambda$ in a neighbourhood of zero.
\end{itemize}
{\bf Remark.} For systems with bosonic heat reservoirs the operator $W$ is not  bounded. Such systems are considered in \cite{MMS}. The development of a general theory for unbounded non-symmetric $W$ is a technically intricate affair, we restrict our attention in this note to systems satisfying condition C (although we give a more general result in Theorem \ref{thm1}).

\subsection{Main results}
\label{subsmainresult}

\subsubsection{Systems close to equilibrium}
\label{subsubclose}

\noindent
We assume that the basic assumptions B1 and B2 are satisfied. The spectral projection onto an eigenvalue $e$ of $L_{\rm S}$ is denoted by $\chi_{L_{\rm S}=e}$ and $P_{\rm R}=|\Phi_{\beta}\rangle\langle\Phi_{\beta}|$ denotes the orthogonal projection onto $\cx\Phi_\beta$. Set $P_e=\chi_{L_{\rm S}=e}\otimes P_{\rm R}$ and $\Pbar_e=\bbbone-P_e$. We allow the case $\dim P_e=\infty$. Denote by $\Lbarlambda^{e}$ the restriction of $L_\lambda$ to $\ran \Pbar_e$, i.e., $\Lbarlambda^{e}:=\Pbar_e L_\lambda \Pbar_e\upharpoonright_{\ran \Pbar_e}$. Consider the following condition:
\begin{itemize}
\item[{\bf A1}$_e$] $P_eI$ and $IP_e$ are bounded operators.
\end{itemize}
If condition A1$_e$ holds then we define the family of bounded operators
\begin{equation}
\Lambda_e(\epsilon) = P_e I\Pbar_e (\overline{L}_0^e-e-\i\epsilon)^{-1} \Pbar_e I P_e,
\label{a8}
\end{equation}
for real $\epsilon\neq 0$. Note that for $\epsilon>0$ we have $\IM \Lambda_e(\epsilon) = \epsilon P_e I\Pbar_e[(\Lbar^e_0-e)^2+\epsilon^2]^{-1}\Pbar_e IP_e \geq 0$, so the numerical range, and hence the spectrum of $\Lambda_e(\epsilon)$, lie in the closed lower complex plane.
If \fer{a8} has a limit as $\epsilon\downarrow 0$ (in the weak sense on a dense domain) then we call this limit the {\it level shift operator} associated to the eigenvalue $e$, and write it as
\begin{equation}
\Lambda_e = P_e I\Pbar_e(\overline{L}_0^e-e-\i 0_+)^{-1}\Pbar_e IP_e.
\label{a9}
\end{equation}
The projection $P_{\rm R}$ has rank one so it is natural to identify $\ran P_e$ with $\ran\chi_{L_{\rm S}=e}$. In this sense we view the operators \fer{a8} and \fer{a9} as operators acting on $\ran \chi_{L_{\rm S}=e}\subset\k\otimes\k$. 
\begin{theorem}
\label{thm3} \ \
\begin{itemize}
\item[(a)] Let $e$ be an eigenvalue of $L_{\rm S}$. {\rm A1}$_e$ holds if and only if {\rm A1}$_{-e}$ holds. If the assumptions {\rm A1}$_{\pm e}$ hold then $J_{\rm S}\Lambda_e(\epsilon) J_{\rm S}=-\Lambda_{-e}(\epsilon)$.  
\item[(b)] Assume A1$_0$. Suppose the spectrum of the Hamiltonian $H$ consists of simple eigenvalues and denote the eigenvectors by $\varphi_i$. Then
$\Lambda_0(\epsilon)=\i\Gamma_0(\epsilon)$, where $\Gamma_0(\epsilon)$ is a selfadjoint positive definite operator on $\ran P_0$ which has real matrix elements in the basis $\{\varphi_i\otimes\varphi_i\}$.  If $\Lambda_0$ exists then the same statements are true for that operator.
\item[(c)] If $H$ has degenerate eigenvalues then neither the real nor the imaginary part of $\Lambda_0(\epsilon)$ vanish, in general. The matrix elements of $\Lambda_0(\epsilon)$ in the basis $\{\varphi_i\otimes\varphi_j\}$ are not purely real nor are they purely imaginary, in general. If $\Lambda_0$ exists then the same statements are true for that operator.
\end{itemize}
\end{theorem}

Part (a) of the theorem shows that $\sigma(\Lambda_e(\epsilon))=-\overline{ \sigma(\Lambda_{-e}(\epsilon))}$. In particular, the spectrum of $\Lambda_0(\epsilon)$ is invariant under reflection at the imaginary axis. Part (b) shows that $\sigma(\Lambda_0(\epsilon))$ lies on the negative imaginary axis if $H$ has simple spectrum. Part (c) says that if $H$ has degenerate spectrum then $\sigma(\Lambda_0(\epsilon))$ can have nonzero real part. In the context of a translation-analytic model of an $N$-level system coupled to a bosonic (or fermionic) heat reservoir \cite{JP1} (see also \cite{MMS}) Theorem \ref{thm3} shows that the set of all resonances is invariant under reflection at the imaginary axis, and that resonances bifurcating from the origin stay on the imaginary axis while wandering into the lower complex plane if $H$ is non-degenerate (see also \fer{*}, \fer{d18.9} and \fer{d19}).

We present a proof of assertions (a) and (b) of Theorem \ref{thm3} in Section \ref{pr1} below.  In Section \ref{degspec} we give examples illustrating statement (c).

Our next result concerns the interplay between the KMS state $\Omega_{\beta,\lambda}$, \fer{a20}, and the level shift operator for $e=0$, $\Lambda_0$. We introduce the following assumptions.
\begin{itemize}
\item[{\bf A2}] $\lambda\mapsto P_0I\Pbar_0(\Lbarlambda-\i\epsilon)^{-1}\Pbar_0$ is continuous at $\lambda=0$  as a map from $\rx$ to the bounded operators on $\h$, for nonzero (small) $\epsilon$. (We write here $\Lbarlambda$ instead of $\Lbarlambda^{0}$.)
\item[{\bf A3}] $\lambda\mapsto \Omega_{\beta, \lambda}$ is differentiable at $\lambda=0$ as a map from $\rx$ to $\h$.
\end{itemize}

\begin{theorem}
\label{thm2}
Assume that Conditions A1$_0$, A2 and A3 hold. Then we have
\begin{equation}
\lim_{\epsilon\rightarrow 0} P_0I\Pbar_0 (\overline{L}_0-\i\epsilon)^{-1} \Pbar_0 IP_0\Omega_{\beta,0} = P_0IP_0\ \partial_\lambda|_{\lambda =0} \Omega_{\beta,\lambda}.
\label{a10}
\end{equation}
In particular, if $\Lambda_0$ exists and if $P_0IP_0=0$ then $\Lambda_0\Psi_\beta=0$.
\end{theorem}

Even though we do not assume that $\Lambda_e(\epsilon)$ converges, equation \fer{a10} shows that $\Lambda_e(\epsilon)\Omega_{\beta,0}$ has a limit as $\epsilon\rightarrow 0$, regardless of the sign of $\epsilon$. By subtracting from \fer{a10} that same equation with $\epsilon$ replaced by $-\epsilon$ we get
\begin{equation}
P_0I\Pbar_0\ \delta(\overline{L}_0)\, \Pbar_0 IP_0\Omega_{\beta,0}=0,
\label{11a}
\end{equation}
and by adding the two equations we obtain
\begin{equation}
P_0I\Pbar_0\ {\rm P.V.}(\overline{L}_0)^{-1}\ \Pbar_0 IP_0\psi_0=P_0IP_0\ \partial_\lambda|_{\lambda=0} \Omega_{\beta,\lambda},
\label{12a}
\end{equation}
where $\delta(x)=\lim_{\epsilon\downarrow 0}\frac 1\pi \frac{\epsilon}{x^2+\epsilon^2}$ is the Dirac delta distribution and ${\rm P.V.} x^{-1}=\lim_{\epsilon\downarrow 0}\frac{x}{x^2+\epsilon^2}$ is the principal value distribution. (The limits are understood in the strong sense.)

{\bf Remarks.\ } 1)\ If $\Lambda_0$ exists then \fer{11a} shows that the Gibbs state $\Psi_\beta$ of the small system belongs to the kernel of ${\rm Im\,}\Lambda_0$, c.f. \fer{a3.9}. If $P_0IP_0=0$, then \fer{a10} implies the second statement of the theorem.

2)\ If $\dim\k<\infty$, $P_0IP_0=0$, and if $\Lambda_0$ exists, a characterization of $\ker (\IM\Lambda_0)$ which implies that $(\IM\Lambda_0)\Psi_\beta=0$ has recently been given in \cite{DF1}.

\begin{theorem}
\label{thm4}
Let $V\in\mm$. Then Conditions B1, B2, A1$_e$ and A2, A3 hold for all eigenvalues $e$ of $L_{\rm S}$. In fact, the maps $\lambda\mapsto \Pbar_0 (\Lbarlambda-\i\epsilon)^{-1}\Pbar_0$ and $\lambda\mapsto\Omega_{\beta,\lambda}$, appearing in Conditions A2 and A3, extend analytically (as $\b(\h)$-valued and $\h$-valued maps) to a complex neighbourhood of $\lambda=0$.
\end{theorem}

\subsubsection{Systems far from equilibrium}
\label{subsubfar}

We assume that condition C is satisfied. The following result is analogous to the one of Theorem \ref{thm2}. Let $P$ be the orthogonal projection onto the kernel of $L_0$, $\Pbar=\bbbone-P$, and denote by $\Lbar_0$ the restriction of $L_0$ to $\ran \Pbar$.

\begin{theorem}
\label{thm5}
Suppose $\lambda\mapsto\psi_\lambda$ is differentiable at $\lambda=0$ as a map from $\rx$ to $\h$ and denote its derivative at zero  by $\psi'_0$. Then we have
\begin{equation}
\lim_{\epsilon\rightarrow 0}PW\Pbar (\Lbar_0-\i\epsilon)^{-1} \Pbar WP\psi_0 = PWP\psi'_0.
\label{n5}
\end{equation}
\end{theorem}
A special case (which is of interest in concrete applications) is given by a reference state  $\psi_\lambda=\psi_0$ which does not depend on $\lambda$, or by interactions satisfying $PWP=0$. In either case \fer{n5} shows that $P\Psi_0\in\ker\Lambda_0$.

\section{Examples}
\label{exsect}

A detailed description of the theory of ideal quantum gases is
given in \cite{BR, MartinRothen, M1, JP}.

\subsection{Reservoirs of thermal Fermions}

The Hilbert space of states of an infinitely extended ideal Fermi
gas which are normal w.r.t. the equilibrium state at inverse
temperature $0<\beta<\infty$ is
\begin{equation}
\h_{\rm R} = \ff_-\otimes\ff_-
\label{a100}
\end{equation}
where $\ff_-=\bigoplus_{n\geq 0}\otimes_{\rm antisym}^n\hh$ is the
antisymmetric Fock space over the one-particle Hilbert space
$\hh=L^2(\r^3,\d^3k)$ (momentum representation). The thermal
creation operators (distributions) are given by
\begin{equation}
a_\beta^*(k) = \sqrt{1-\mu_\beta} \, a^*(k) \otimes\bbbone_{\ff_-}
+ (-1)^N \otimes \sqrt{\mu_\beta} \, a(k),\ \ \ k\in{\mathbb R}^3,
\label{a101}
\end{equation}
whith $\mu_\beta(k)=(1+\e^{\beta\omega(k)})^{-1}$ and where
\begin{equation}
k\mapsto \omega(k)\geq 0
\label{a101.1}
\end{equation}
is the dispersion relation of the fermions considered. A typical
example are non-relativistic fermions for which $\omega(k)=|k|^2$.
The $a$ and $a^*$ on the r.h.s. of \fer{a101} are the ordinary
fermionic Fock annihilation and creation operators which satisfy
the canonical anti-commutation relations
\begin{equation}
\{a(k),a^*(l)\}=\delta(k-l).
\label{a102}
\end{equation}
The number operator $N$ in \fer{a101} is given by $N=\int_{\r^3}
a^*(k)a(k)\,\d^3 k $. Relations \fer{a100} and \fer{a101}
constitute the so-called {\it Araki-Wyss representation} of the
canonical anti-commutation relations \cite{AWyss}. Smeared-out
creation and annihilation operators are defined by
$a^*(f)=\int_{\r^3} f(k)a^*(k)\d^3 k$ and $a(f) = \int_{\r^3}
\overline{f}(k)a(k)\d^3 k$, for $f\in\hh$, and where $\overline f$
stands for the complex conjugate. One shows that the fermionic
creation and annihilation operators are bounded, satisfying
$\|a^\#(f)\|=\|f\|_{\hh}$. The von Neumann algebra $\mm_{\rm R}$
is generated by the thermal creation and annihilation operators
$\{a_\beta^\#(f)\ |\ f\in\hh\}$. The vector
\begin{equation}
\Phi_\beta =\Omega\otimes\Omega\in\h_{\rm R},
\label{a105}
\end{equation}
where $\Omega$ is the (Fock) vacuum vector in $\ff_-$, represents the KMS state w.r.t. the dynamics $t\mapsto \e^{\i tL_{\rm R}}a^\#_\beta(f) \e^{-\i tL_{\rm R}}=a^\#_\beta(\e^{\i t\omega}f)$, where
\begin{equation}
L_{\rm R} = \d\Gamma(\omega)\otimes\bbbone_{\ff_-} - \bbbone_{\ff_-}\otimes\d\Gamma(\omega),
\label{a106}
\end{equation}
and $\d\Gamma(\omega)=\int_{\r^3}\omega(k) a^*(k)a(k)\, \d^3k$ is the second quantization of multiplication by $\omega(k)$. The action of the modular conjugation $J_{\rm R}$ on creation operators is
\begin{equation}
J_{\rm R} a_\beta^*(f) J_{\rm R} = \big[ (-1)^N\otimes a^*(\sqrt{1-\mu_\beta}\, \overline f) + a(\sqrt{\mu_\beta}\, \overline{f})\otimes\bbbone_{\ff_-}\big] (-1)^N\otimes(-1)^N.
\label{a106.1}
\end{equation}

Now we describe the interaction with the small system. The latter is described at the beginning of Section \ref{subs1}. For $m,n\geq 0$, $m+n\geq 1$, let  $h_{m,n}$ be maps from $\r^{3m}\times\r^{3n}$ to the bounded operators on $\k$. For simplicity of exposition, we assume that those maps are continuous and have compact support. We define $\bk^{(m)}=(k_1,\ldots,k_m)\in\r^{3m}$ and put
\begin{equation}
a^*_\beta(\bk^{(m)})= a^*_\beta(k_1)\cdots a^*_\beta(k_m),
\label{106.2}
\end{equation}
and similarly for $a_\beta(\bl^{(n)})$.  Set
\begin{equation}
v_{m,n}= \int_{\r^{3m}\times\r^{3n}} \d\bk^{(m)} \d\bl^{(n)}\  h_{m,n}(\bk^{(m)},\bl^{(n)})\otimes  a^*_\beta(\bk^{(m)}) a_\beta(\bl^{(n)}) +\mbox{\ adjoint}.
\label{a103}
\end{equation}
It is well known that
\begin{equation}
V_{m,n}=\pi(v_{m,n})\in\mm,
\label{a103.3}
\end{equation}
see \fer{a.3.91}. The following result follows from Theorem \ref{thm4}.
\begin{theorem}
\label{fermionsthm}
Suppose $V$ is a linear combination of terms \fer{a103.3}. Then all conditions B1, B2, A1$_e$, A2, A3 hold. In particular, the conclusions of Theorems \ref{thm3} and \ref{thm2} are valid.
\end{theorem}

\subsection{Reservoirs of thermal Bosons}
The Hilbert space describing states of an infinitely extended ideal Bose gas which are normal to the equilibrium state at inverse temperature $0<\beta<\infty$ (without Bose-Einstein condensate) is
\begin{equation}
\h_{\rm R}=\Fplus \otimes\Fplus
\label{a60}
\end{equation}
where $\Fplus=\bigoplus_{n\geq 0}\otimes_{\rm sym}^n \hh$ is the
symmetric Fock space over the one-particle Hilbert space
$\hh=L^2(\r^3,\d^3k)$ (momentum representation). The thermal
creation operators (distributions) are given by
\begin{equation}
a_\beta^*(k) = \sqrt{1+\rho_\beta} \, a^*(k) \otimes\bbbone_{\Fplus} + \bbbone_{\Fplus}\otimes \sqrt{\rho_\beta} \, a(k),\ \ \ k\in{\mathbb R}^3,
\label{a61}
\end{equation}
where $\rho_\beta(k)=(\e^{\beta\omega(k)}-1)^{-1}$ is the momentum density distribution, and
\begin{equation}
k\mapsto\omega(k)\geq 0
\label{a61.1}
\end{equation}
is the dispersion relation. We shall consider for simplicity of the exposition $\omega$'s such that $\rho_\beta(k)$ is locally integrable in $\r^3$. A typical example is $\omega(k)=|k|$ (massless relativistic Bosons). The $a$ and $a^*$ on the r.h.s. of \fer{a61} are the ordinary Fock annihilation and creation operators which satisfy the canonical commutation relations
\begin{equation}
[a(k),a^*(l)]=\delta(k-l).
\label{a62}
\end{equation}
The representation \fer{a60}, \fer{a61} is the so-called {\it Araki-Woods representation} of the canonical commutation relations \cite{AW}.
Smeared-out creation and annihilation operators are defined by $a^*(f)=\int_{\r^3} f(k)a^*(k)\d^3 k$ and $a(f) = \int_{\r^3} \overline{f}(k)a(k)\d^3 k$, for $f\in\hh$, where $\overline f$ stands for the complex conjugate.
The von Neumann algebra $\mm_{\rm R}$ is generated by the Weyl operators $W_\beta(f)=\e^{\i\phi_\beta(f)}$, $f\in\hh$, where
\begin{equation}
\phi_\beta(f)=\frac{1}{\sqrt 2}\big(a_\beta^*(f)+a_\beta(f)\big)
\label{a62.1}
\end{equation}
is the selfadjoint field operator. The unbounded operators $a_\beta^*(f)$, $a_\beta(f)$ and $\phi_\beta(f)$ are affiliated with $\mm_{\rm R}$. The vector
\begin{equation}
\Phi_\beta =\Omega\otimes\Omega\in\h_{\rm R},
\label{a65}
\end{equation}
where $\Omega$ is the (Fock) vacuum vector in $\ff_+$, represents the KMS state w.r.t. the dynamics $t\mapsto \e^{\i tL_{\rm R}}W_\beta(f)\e^{-\i tL_{\rm R}}=W_\beta(\e^{\i t\omega}f)$, where
\begin{equation}
L_{\rm R} = \d\Gamma(\omega)\otimes\bbbone_{\ff_+} - \bbbone_{\ff_+}\otimes\d\Gamma(\omega),
\label{a66}
\end{equation}
and $\d\Gamma(\omega)=\int_{\r^3}\omega(k)a^*(k)a(k)\, \d^3k$ is the second quantization of multiplication by $\omega(k)$. The modular conjugation $J_{\rm R}$ acts on creation operators as
\begin{equation}
J_{\rm R} a_\beta^*(f) J_{\rm R} = \bbbone_{\ff_+}\otimes a^*(\sqrt{1+\rho_\beta}\, \overline f) + a(\sqrt{\rho_\beta}\, \overline{f})\otimes\bbbone_{\ff_+}.
\label{a66.1}
\end{equation}
We now turn to the interaction between this reservoir and the small system (which is described at the beginning of Section \ref{subs1}). It is given in a similar way as for Fermions, but, for technical reasons, it has to be restricted to at most quadratic expressions in the thermal creation and annihilation operators (``two--body interactions''). Given $h_1$ and $h_2$, continuous maps from $\r^3$ and from $\r^3\times\r^3$ to the bounded operators on $\k$ respectively, s.t. $h_2(k,l)=h_2(l,k)^*$, we set
\begin{equation}
v= \int_{\r^3}\big[ h_1(k)\otimes a_\beta^*(k)  + h_1(k)^*\otimes a_\beta(k)\big]\d^3k + \int_{\r^3\times\r^3} h_2(k,l)\otimes a_\beta^*(k)a_\beta(l) \ \d^3k\, \d^3l.
\label{a63}
\end{equation}
For the purpose of exposition we assume that $h_{1,2}$ have compact support. We introduce
\begin{eqnarray}
C(s) &=& \int_{\r^3}
\e^{2s\omega(k)}\|\e^{-sH}h_1(k)\e^{sH}\|^2
\d^3k \nonumber\\
&& +\int_{\r^3\times\r^3} \e^{-2s(\omega(k)-\omega(l))} \|\e^{-sH}h_2(k,l)\e^{sH}\|^2\ d^3k\, \d^3l,
\label{a64}
\end{eqnarray}
which measures the regularity of the integral kernels $h_{1,2}$ relative to the Hamiltonian $H$.
\begin{theorem}
\label{thmbosons1}
Suppose the functions $h_1$ and $h_2$ are such that
\begin{equation}
\sup_{-\beta/2 \leq s \leq\beta/2} C(s) =C<\infty\label{a70}.
\end{equation}
There is a constant $\lambda_0$ (depending on $\beta$) s.t. if $|\lambda|<\lambda_0$ then all the conditions B1, B2, A1--A3 are met.
\end{theorem}

\noindent
{\bf Remarks.\ } 1) If $\dim \k<\infty$ then \fer{a70} is satisfied.

2) We can replace the condition of $h_{1,2}$ having compact support by a suitable weaker decay condition. Moreover, assuming $h_{1,2}$ to satisfy certain regularity properties at $k,l=0$ we can treat discontinuous $h_{1,2}$.

3) One may add to $v$, \fer{a63}, integrals involving $a_\beta(k)a_\beta(l)$, $a_\beta(k)a^*_\beta(l)$, $a^*_\beta(k)a^*_\beta(l)$.

4) If $h_2=0$ then the theorem holds for all $\lambda\in\rx$ (see the remark at the end of the proof of Theorem \ref{thmbosons1}, Section \ref{sectproofthmbosons1}).

\subsection{Systems with several reservoirs}

We consider $R\geq 2$ fermionic reservoirs coupled through a small system. The Hilbert space is given by $\h=\h_{\rm S}\otimes \h_{{\rm R}_1}\otimes \cdots \otimes \h_{{\rm R}_R}$, with an uncoupled dynamics generated by $L_0=L_{\rm S} + L_{{\rm R}_1}+ \cdots +L_{{\rm R}_R}$. We may choose the reference state to be a product of equilibrium states, $\psi_0=\Psi_{\beta_{\rm S}}\otimes\Phi_{\beta_1}\otimes\cdots \otimes \Phi_{\beta_R}$. Denote by $J_{{\rm R}_i,{\rm S}}$ and $\Delta_{{\rm R}_i,{\rm S}}$ the modular conjugations and the modular operators associated to the pairs $(\mm_{{\rm R}_i},\Phi_{\beta_i})$ and $(\mm_{\rm S},\Psi_{\beta_{\rm S}})$, respectively. Take $V_1,\ldots, V_R$ to be linear combinations of the form \fer{a103.3}, as in Theorem \ref{fermionsthm}, representing the couplings to the reservoirs, and suppose that
\begin{equation*}
(\Delta_{\rm S}^{1/2}\otimes\Delta_{{\rm R}_i}^{1/2}) \ V_i \ (\Delta_{\rm S}^{-1/2}\otimes\Delta_{{\rm R}_i}^{-1/2}) \in \mm_{\rm S}\otimes\mm_{\rm R}.
\end{equation*}
We set
\begin{equation}
W = \sum_i \left\{ V_i -  \big( J_{\rm S} \Delta_{\rm S}^{1/2}\otimes J_{{\rm R}_i}\Delta_{{\rm R}_i}^{1/2}\big) \ V_i \  \big( J_{\rm S} \Delta_{\rm S}^{1/2}\otimes J_{{\rm R}_i}\Delta_{{\rm R}_i}^{1/2}\big)   \right\}.
\end{equation}
It is easily verified that $W\psi_0=0$ and thus $K_\lambda\psi_0 = (L_0+\lambda W)\psi_0=0$. Since the second part in the sum belongs to the commutant $\mm'$, $K_\lambda$ defines the same dynamics on $\mm$ as does $L_0+\lambda (V_1+\cdots +V_R)$. Thus condition C is satisfied and Theorem \ref{thm5} applies.

\subsection{Klein--Gordon field for accelerated observer}

Let $x=(x^0,\ux)$ be a point in Minkowski space-time $\r\times\r^3$ (with metric signature $(+,-,-,-)$). The field operator satisfying the Klein--Gordon equation
\begin{equation}
(\square+m^2)\varphi(x)=0,
\label{b1}
\end{equation}
where $\square=\partial_{x^0}^2-\Delta$, $\Delta$ is the Laplacian, $m\geq 0$, is given by
\begin{equation}
\varphi(x)=\int_{\r^3}\frac{\d\uk}{\sqrt{2\omega(\uk)}}\big[ \e^{\i \uk\,\ux -\i\omega(\uk) x^0}a(\uk) +\e^{-\i \uk\,\ux +\i\omega(\uk) x^0}a^*(\uk)\big].
\label{b2}
\end{equation}
The $a^\#(\uk)$ in \fer{b2} are the usual bosonic creation and annihilation operators satisfying \fer{a62}, and
\begin{equation}
\omega(\uk) = \big(\uk^2+m^2\big)^{1/2}.
\label{b3}
\end{equation}
Let ${\cal S}(\r^4;\r)$ and ${\cal S}(\r^4;\cx)$ denote the real valued and the complex valued Schwartz functions on $\r^4$, respectively. For $f\in{\cal S}(\r^4;\cx)$ we define the smeared field operators, acting on bosonic Fock space $\ff_+$ (see after \fer{a60}), by
\begin{equation}
\varphi[f] = \int_{\r^4} f(x)\varphi(x)\,\d^4x.
\label{b4}
\end{equation}
The adjoint of $\varphi[f]$ is $\varphi[\,\overline f\,]$, where $\overline f$ stands for the complex conjugate of $f$. For $f\in{\cal S}(\r^4,\r)$ we define the unitary Weyl operator $W[f]=\e^{\i\varphi[f]}$.

We introduce the ``right wedge'' ${\cal W}_{\rm R}=\{x\in\r^4\, |\  |x^0|<x^1 \}$.
Let $\mm_{\rm R}\subset {\cal B}(\ff_+)$ be the von Neumann algebra generated by
\begin{equation}
\{W[f]\, |\, f\in{\cal S}(\r^4,\r), \,\supp(f)\subset{\cal W}_{\rm R}\},
\label{b5}
\end{equation}
where $\supp(f)$ denotes the support of $f$. Elements of $\mm_{\rm R}$ are interpreted to be those observables which can be measured in the space-time region ${\cal W}_{\rm R}$.

It is well known (Reeh--Schlieder) that the vacuum vector $\Omega_{\rm R}\in\ff_+$ is cyclic for the $*$algebra of all polynomials in $\varphi[f]$,  $f\in{\cal S}(\r^4;\cx)$. We may thus define an antiunitary involution $J_{\rm R}$ on $\ff_+$ by
\begin{equation}
J_{\rm R}\varphi[f(x)]J_{\rm R}=\varphi[\, \overline f(-x^0,-x^1,x^2,x^3)],
\label{b6}
\end{equation}
for all $f\in{\cal S}(\r^4,\cx)$. Relativistic boosts in the $x^1$-direction are given by
\begin{equation*}
B_\tau: (x^0,x^1,x^2,x^3)\mapsto (x^0\cosh\tau+x^1\sinh\tau, x^0\sinh\tau+x^1\cosh\tau,x^2,x^3),
\label{b7}
\end{equation*}
for $\tau\in\rx$. The action of $B_\tau$ lifts to Fock space according to
\begin{equation}
\varphi[f\circ B_{-\tau}] = \e^{\i\tau L_{\rm R}}\varphi[f]\e^{-\i\tau L_{\rm R}},
\label{b8}
\end{equation}
where $L_{\rm R}$ is the selfadjoint Liouville operator on $\ff_+$ given by 
\begin{equation}
L_{\rm R} = \d\Gamma\big([-\Delta+m^2]^{1/4} \,x^1\, [-\Delta+m^2]^{1/4}\big).
\label{b8.1}
\end{equation}
The map $B_\tau$ leaves the wedge ${\cal W}_{\rm R}$ invariant so
\begin{equation}
\alpha_{\rm R}^\tau(A) = \e^{\i \tau L_{\rm R}} A \e^{-\i \tau L_{\rm R}}
\label{b11}
\end{equation}
defines a group of $*$automorphisms of $\mm_{\rm R}$.
\begin{theorem}[\cite{BW}]
\label{bisognanowichmann}
The state on $\mm_{\rm R}$ determined by the vacuum vector $\Omega_{\rm R}\in\ff_+$ is a $(2\pi,\alpha_{\rm R}^\tau)$-KMS state. The modular operator associated to $(\mm_{\rm R},\Omega_{\rm R})$ is $\Delta_{\rm R}=\e^{-2\pi L_{\rm R}}$ and the modular conjugation is $J_{\rm R}$.
\end{theorem}
An observer accelerating in the $x^1$-direction with constant acceleration $a>0$, with position $(1/a,x^2,x^3)$ at the instant $x^0=0$, describes the curve \cite{Marder}
\begin{equation}
\left[
\begin{array}{c}
x^0 \\ x^1 \\ x^2\\ x^3
\end{array}
\right]
=
B_{\tau(x^0)}
\left[
\begin{array}{c}
0\\ 1/a \\ x^2\\ x^3
\end{array}
\right],
\label{b26}
\end{equation}
where $\e^{\tau(x^0)} = ax^0 +\sqrt{a^2(x^0)^2+1}$.
The $(x')^1$-axis of the accelerated observer's instantaneous rest frame is given by the half line emanating from the origin and passing through the point \fer{b26}, in the coordinate system $(x^0,\ldots,x^3)$. The curve $B_{\tau(x^0)}[0 \ 1/b \ x^2\  x^3]^T$ traces the point which is perceived by the accelerated observer to be at rest, lying on the $(x')^1$-axis at $\ln(b/a)/a$ (in a natural coordinate system) \cite{Marder}. 

According to \fer{b8} and \fer{b26}, the field at the point of the accelerated observer at time $x^0$ is given by
\begin{equation}
\varphi(x^0,\ux) = \e^{\i\tau(x^0) L_{\rm R}} \varphi(0,1/a,x^2,x^3) \e^{-\i\tau(x^0)L_{\rm R}}.
\label{b28}
\end{equation}
Similarly, the field at the moment $x^0$, in a space point $\ux$ which is at rest as seen by the accelerated observer, with spatial coordinates $(\delta',x^2,x^3)$ in the instantaneous rest frame, is given by
\begin{equation}
\varphi(x^0,\ux) = \e^{\i\tau(x^0) L_{\rm R}} \varphi(0,\e^{\delta'}\!\!/a,x^2,x^3) \e^{-\i\tau(x^0)L_{\rm R}}.
\label{b29}
\end{equation}
The proper time (instantaneous time) on a trajectory of an accelerated observer in $x^1$-direction with constant acceleration $a>0$ is $\tau_a=\tau/a$, \cite{Marder,S}. Theorem \ref{bisognanowichmann} thus means that the field, in the vacuum state $\Omega_{\rm R}$, is perceived by the accelerated observer as a {\it thermal field} at temperature $\beta=2\pi/a$ (``Unruh Effect'').

To measure this temperature concretely, the accelerated observer carries a thermometer, which we model by a small system, as described at the beginning of Section \ref{subs1}. 
The interaction between the thermometer and the ambient field 
is given by the operator \cite{UW,DM}
\begin{equation}
V= G\otimes \bbbone_{\k}\otimes \int_{{\cal W}_{\rm R}}\d^4x \ \delta(x^0)\ \rho(\ux)\, \varphi(x),
\label{b21}
\end{equation}
or by a sum of such operators, where the integral is effectively taken over the time-slice $x^0=0$. Here, $G$ is a bounded selfadjoint operator on $\k$ and $\rho$ is a smooth function compactly supported in a small neighbourhood of the position of the accelerated observer at time zero, $(1/a,0,0)$.

\begin{theorem}
\label{thmaccobs}
Suppose that $\sup_{0\leq s\leq \pi}\|\e^{-s H}G\e^{s H}\|<\infty$. Then all the conditions B1, B2, A1--A3 are met for arbitrary $\lambda\in\rx$.
\end{theorem}

We give a proof of this result in Section \ref{sectproofthmaccobs}. Theorem \ref{thmaccobs} is useful in the mathematical analysis of the Unruh Effect, \cite{DM}.

\section{Hamiltonians with degenerate spectrum}
\label{degspec}

We consider a small system with $\dim\k<\infty$ coupled to a reservoir of relativistic massless Bosons (dispersion relation $\omega(k)=|k|$) at temperature $\beta>0$. The interaction is
\begin{equation}
V =  G\otimes\bbbone_{\cx^2}\otimes \phi_\beta(g),
\label{d1}
\end{equation}
where $G$ is a selfadjoint operator on $\k$, $g\in L^2(\r^3,\d^3k)$, and where the thermal field operator $\phi_\beta(g)$ is defined in \fer{a62.1}.
\begin{lemma}
\label{degspeclemma}
Assume that $g$ has compact support and is continuous, except possibly at $|k|=0$, and suppose that
\begin{equation}
\lim_{|k|\rightarrow 0} \frac{|g(k)|}{|k|^p}=\gamma,
\label{d2}
\end{equation}
for some $p>-1$ and some $\gamma\geq 0$. Then the system introduced in this paragraph, with interaction \fer{d1}, satisfies all conditions B1, B2, A1$_e$, A2 and A3, for any value of the coupling constant $\lambda$.
\end{lemma}
We do not optimize the ultraviolet behaviour of $g$ here. A proof of this result is easily obtained by repeating the proof of Theorem \ref{thmbosons1}.

\begin{theorem}
\label{thmdegspec}
Let $\k=\cx^2$ and $H={\rm diag}(e,e)$, for some $e\in\r$. We have $\dim P_0=4$.
Set $\xi = \|g/\sqrt\omega\|^2/2$ and $\eta=\frac{2\pi^2\gamma^2}{3\beta}$.
\begin{itemize}
\item[1.] If $-1<p<-1/2$ then the level shift operator $\Lambda_0$ does not exist.
\item[2.] If $p=-1/2$ then the level shift operator $\Lambda_0$ exists and is given by
\begin{equation}
\Lambda_0 = (\xi+\i\eta) G^2\otimes\bbbone_{\cx^2} -(\xi-\i\eta)\bbbone_{\cx^2}\otimes\cc G^2\cc -2\i\eta G\otimes\cc G\cc,
\label{d4}
\end{equation}
where $\cc$ is the antiunitary map taking the complex conjugate of coordinates in the canonical basis of $\cx^2$.
\item[3.] If $p>-1/2$ then $\Lambda_0$ exists and is given by \fer{d4} with $\eta=0$.
\end{itemize}
\end{theorem}
This result is obtained by explicit calculation, see Section \ref{proofthmdegspec}. Relation \fer{d4} gives
\begin{eqnarray}
\RE\Lambda_0 &=& \xi \big( G^2\otimes\bbbone_{\cx^2} -\bbbone_{\cx^2}\otimes\cc G^2\cc\big)\\
\IM\Lambda_0 &=&\eta\big( G\otimes\bbbone_{\cx^2} - \bbbone_{\cx^2}\otimes\cc G\cc\big)^2\geq 0.
\end{eqnarray}
For $p>-1/2$ we have $\IM\Lambda_0=0$. The particular structure of $\Lambda_0$ allows us to find its spectrum easily. Let $\alpha_1$, $\alpha_2$ be the (real) eigenvalues of $G$, with associated eigenvectors $\chi_1$, $\chi_2$. Then the eigenvectors of $\Lambda_0$ are $\chi_i\otimes\chi_j$, $i,j\in\{ 1,2\}$, and the spectrum of $\Lambda_0$ is
\begin{equation}
\sigma(\Lambda_0) = \{0,0,z, -\overline z\},
\label{d5}
\end{equation}
 where $z=(\alpha_1-\alpha_2) [\xi(\alpha_1+\alpha_2)+\i\eta(\alpha_1-\alpha_2)]$. These observations prove the assertions of Theorem \ref{thm3}, part (c).

\medskip

A physically more interesting model is given by a small system with a non-degenerate ground state and two degenerate excited states. We have $\k=\cx^3$, $H={\rm diag}(e,f,f)$ with $\Delta=f-e>0$ and $\dim P_0=5$. Take a coupling of the form \fer{d1}, where
\begin{equation}
G=
\left[
\begin{array}{ccc}
0 & a & b \\
a & 0 & c \\
b & c & 0
\end{array}
\right], \mbox{\ \ \ with $a,b,c\in\rx$}.
\label{d10}
\end{equation}

\begin{theorem}
\label{degthm2} Consider the interaction \fer{d1} with $G$ as in
\fer{d10} and assume $g(k)$ satisfies \fer{d2} with $p>-1/2$. Then
$\Lambda_0$ exists and has spectrum
\begin{equation}
\sigma(\Lambda_0)=\Big\{0,0,is(a^2+b^2)\coth(\beta\Delta/2),
(a^2+b^2)(\pm\alpha+\frac{\i
s}{2}\frac{e^{\beta\Delta}}{\e^{\beta\Delta}-1})\Big\},
\end{equation}
where $s=\pi \Delta^2 \int_{S^2}\d\sigma |g(\Delta,\sigma)|^2$ and
$\alpha = {\rm
P.V.}\scalprod{g}{\frac{1+\rho_\beta}{\Delta-\omega}g}$ (principal
value).
\end{theorem}

{\bf Remark.\ } The operator $\Lambda_0$ is independent of $c$,
see \fer{d11}. This comes from the fact that $c$ governs
transitions between the two degenerate energy levels (Bohr
frequency zero) and $g$ has a ``mild'' singularity at the origin
($p>-1/2$). We understand this independence of the level shift
operator of $c$ as the origin of the degeneracy of its kernel (the
coupling is ``not effective'' since one may choose $c=0$).
$\Lambda_0$ exists and depends on $c$ in case $p=-1/2$, c.f. after
\fer{d12}.

The violation of the Fermi Golden Rule Condition ($\dim\ker
\Lambda_0>1$) calls for a more detailed examination of the time
ergodic properties of the system. For the purpose of exposition,
we consider a translation analytic \cite{JP1} model in what
follows. The Feshbach map applied to the spectrally translated
$L_\lambda(\tau)$ is \cite{BFS1}
\begin{equation}
F(L_\lambda(\tau)) = -\lambda^2 P_0I\Pbar_0
(\Lbar_\lambda(\tau))^{-1}\Pbar_0 IP_0 = -\lambda^2
\lim_{\epsilon\downarrow 0}
P_0I\Pbar_0(\Lbar_\lambda-\i\epsilon)^{-1}\Pbar_0 IP_0,
\label{d18.9}
\end{equation}
where we write $\Lbar$ instead of $\Lbar^0$. We take the
convention that the spectral deformation is implemented in such a
way that the the spectrum of $L_\lambda$ gets pushed into the
lower complex plane for the range of $\tau$ considered (hence the
sign $\epsilon\downarrow 0$). We expand
\begin{equation}
P_0I\Pbar_0(\Lbar_\lambda-\i\epsilon)^{-1}\Pbar_0 IP_0=
\Lambda_0(\epsilon) + \lambda^2 \Lambda'_0 (\epsilon)+
O(\lambda^4), \label{d19}
\end{equation}
see also \fer{a8} and \fer{*}, with  a remainder $O(\lambda^4)$
 uniform in $\epsilon$ for small $\epsilon$. The following
result describes the spectrum of $\Lambda_0+\lambda^2\Lambda_0'$,
where $\Lambda_0'=\lim_{\epsilon\downarrow
0}\Lambda_0'(\epsilon)$.

\begin{proposition}
\label{degprop} Consider the interaction matrix $G$, \fer{d10},
with $a\neq 0$, $b=0$. The operator $\Lambda'_0$ exists and the
spectrum of $\Lambda_0+\lambda^2\Lambda_0'$ in a neighbourhood of
order $\lambda^2$ around the origin consists of two eigenvalues,
one at the origin and another one, $z$, which satisfies
\begin{equation}
\IM z=a^2c^2 \xi_1(\Delta)+ c^4\xi_2, \label{d20}
\end{equation}
where $\xi_1(\Delta),\xi_2>0$. $\xi_1(\Delta)$ is of order
$\Delta^2$ for small $\Delta$, while $\xi_2$ does not depend on
$\Delta$.
\end{proposition}
The operator $\Lambda_0'$ exists also when $b\neq 0$ and it is
possible (albeit much longer) to calculate the spectrum of
$\Lambda_0+\lambda^2\Lambda'_0$ in that case. We give explicit
expressions for $\xi_1$, $\xi_2$ in equations \fer{xi1},
\fer{xi2}.

It follows from \fer{d20} that the degeneracy of the eigenvalue
zero of $\Lambda_0$ is lifted by the perturbation
$\lambda^2\Lambda_0'$. By the isospectrality of the Feshbach map
one concludes from \fer{d18.9} and \fer{d19} that $\ker L_\lambda$
is simple, provided $\lambda\neq 0$ is small enough. Moreover, it
is shown in a standard way that all non-zero resonances of
$L_\lambda$ have strictly negative imaginary part. This implies
the property of return to equilibrium with exponentially fast
convergence, proportional to $\e^{-\xi\lambda^4t}$, for some
$\xi>0$.

\section{Proofs}
\label{proofsection}

\subsection{Proof of Theorem \ref{thm3}}
\label{pr1}

We realize the Hilbert space of the Gibbs state $\Psi_\beta$ of the small system by $\h_{\rm S}=\k\otimes\k$. Let $\{E_i\}$, $E_0\leq E_1\leq\cdots$, denote the spectrum of $H$ and let $\varphi_i$ the normalized eigenvector associated to $E_i$. The $\varphi_i$ form an orthonormal basis of $\k$.  Then
\begin{equation}
\Psi_\beta= Z_\beta^{-1} \sum_{i\geq 0} \e^{-\beta E_i/2} \varphi_i\otimes\varphi_i,
\label{a12}
\end{equation}
where $Z_\beta$ is a normalization factor. It is manifest that $L_{\rm S}\Psi_\beta=0$, c.f. \fer{a13}. The modular conjugation $J_{\rm S}$ has the explicit action
\begin{equation}
J_{\rm S}\psi\otimes\phi = \phi\otimes \cc\psi\cc
\label{a14}
\end{equation}
where $\cc$ is the antilinear operator whose effect is to take the complex conjugate of coordinates of vectors in $\k$, relative to the basis $\{\varphi_i\}$. As is easily verified, we have the relations $J_{\rm S}\Psi_\beta=\Psi_\beta$ and $J_{\rm S} L_{\rm S}J_{\rm S}=-L_{\rm S}$ (c.f. \fer{a13}). The spectrum of $L_{\rm S}$ consists of the eigenvalues $\{E_{i,j}:=E_i-E_j\ |\ i,j\geq 0\}$.

Let $e$ be an eigenvalue of $L_{\rm S}$. A basis for $\ran P_e$ is given by
\begin{equation}
\{ \varphi_i\otimes\varphi_j\otimes\Phi_\beta\ |\ \mbox{$(i,j)$ s.t. $E_i-E_j=e$}\}.
\end{equation}
According to the terminology introduced in \cite{BFS}, Appendix B, if $e$ is an eigenvalue of $L_{\rm S}$, then $\{e\}$ is called a {\it nondegenerate set} if and only if the following holds whenever $i$ is s.t. $E_i-e$ lies in the spectrum of $H$: if $E_i-E_j=e$ and $E_i-E_{j'}=e$ then $j=j'$. {\it It should be noted that $\{0\}$ is a nondegenerate set (in the sense of \cite{BFS}) if and only if all $E_i$ are distinct, i.e., if and only if $H$ has simple spectrum}.
If the spectrum of $H$ is simple then a basis of $\ran P_0$ is given by $\{\varphi_i\otimes\varphi_i\otimes\Phi_\beta\}$.

\bigskip
\noindent
{\bf Proof of Theorem \ref{thm3}.\ }(a) Let $p_i=|\varphi_i\rangle\langle \varphi_i|$ denote the orthogonal projection onto $\cx \varphi_i$. We have
\begin{equation}
P_e=\sum_{E_{i,j}=e} p_i\otimes p_j\otimes P_{\rm R}.
\label{a30}
\end{equation}
Taking into account $J_{\rm S}\varphi_i\otimes\varphi_j=\varphi_j\otimes\varphi_i$, $J_{\rm R}\Phi_\beta=\Phi_\beta$ we obtain for any $\chi\in\h$
\begin{eqnarray}
JP_eJ\chi &=& \sum_{E_{i,j}=e}  \overline{\scalprod{\varphi_i\otimes\varphi_j\otimes\Phi_\beta}{J\chi}}\ \varphi_j\otimes\varphi_i\otimes\Phi_\beta \nonumber\\
&=& \sum_{E_{i,j}=e}  \scalprod{\varphi_j\otimes\varphi_i\otimes\Phi_\beta}{\chi}\ \varphi_j\otimes\varphi_i\otimes\Phi_\beta.
\label{a31}
\end{eqnarray}
To arrive at the second equality we use that $J$ is antiunitary. The equivalence $E_{i,j}=e \Longleftrightarrow E_{j,i}=-e$ and \fer{a31} show that
\begin{equation}
J P_eJ = P_{-e}.
\label{a32}
\end{equation}
Since $JL_0J=-L_0$, $J^2=1$ and $J$ is antilinear, we have
\begin{equation}
J \Pbar_e (\overline{L}_0^e-e-\i\epsilon)^{-1}\Pbar_e J = -\Pbar_{-e} (\overline{L}_0^{-e}+e-\i\epsilon)^{-1}\Pbar_{-e}.
\label{a33}
\end{equation}
Assume that A1$_{e}$ holds. Using \fer{a32}, \fer{a33} and  $JIJ=-I$, c.f. \fer{a5}, we obtain
\begin{equation}
J\Lambda_e(\epsilon) J = -P_{-e}I\Pbar_{-e}(\overline{L}_0^{-e}+e-\i\epsilon)^{-1}\Pbar_{-e}IP_{-e} = -\Lambda_{-e}(\epsilon).
\label{a34}
\end{equation}
This shows assertion (a) of Theorem \ref{thm3}.

\medskip
\noindent
(b) For $\epsilon>0$ we have the representation
\begin{equation}
2\RE (\overline{L}_0^{e}-e-\i\epsilon)^{-1} =\i\int_\rx\d t \ \e^{-\epsilon|t|}\sgn(t) \e^{-\i t(\overline{L}_0^e-e)}
\label{a35}
\end{equation}
where we set $\RE A=\frac 12 (A+A^*)$ for an operator $A$, and $\sgn$ takes the value $1$ for arguments $t\geq 0$ and the value $-1$ for $t<0$. We thus obtain
\begin{equation}
2\RE\Lambda_e(\epsilon) =\i\int_\rx\d t\ \e^{-\epsilon|t|} \sgn(t) \left[ P_e I \e^{-\i t(L_0-e)} I P_e - P_eIP_eIP_e\right],
\label{a36}
\end{equation}
where we use $\Pbar_e=\bbbone-P_e$. The second term on the r.h.s. of \fer{a36} vanishes because the integrand is an odd function. Using $P_e(L_0-e)=0$ we arrive at
\begin{equation}
2\RE\Lambda_e(\epsilon) =\i\int_\rx\d t\ \e^{-\epsilon|t|} \sgn(t) P_e \e^{\i t L_0} I \e^{-\i t L_0} I  P_e.
\label{a37}
\end{equation}
The decomposition $I=V-JVJ$, \fer{a5}, leads to four terms in the integral of \fer{a37}. The contribution of the two ``cross terms'' where both $V$ and $JVJ$ occur is
\begin{equation}
-P_e \e^{\i tL_0}JVJ \e^{-\i tL_0}VP_e - P_e \e^{\i tL_0} V\e^{-\i tL_0} JVJ P_e.
\label{a40}
\end{equation}
The following little result is proven below.
\begin{lemma}
\label{lemma1}
Let $M$ be a selfadjoint operator affiliated with $\mm$, s.t. $M P_e$  and $JMJ P_e$ are bounded operators. Then
\begin{equation}
P_e \ \e^{\i tL_0}M\e^{-\i tL_0} \ JMJ \ P_e= P_e \ JMJ \ \e^{\i tL_0}M\e^{-\i tL_0}\ P_e,
\label{a38}
\end{equation}
for all $t\in\rx$.
\end{lemma}
Lemma \ref{lemma1} shows that $\e^{\i tL_0} V\e^{-\i tL_0}$ in the second term of \fer{a40} can be commuted through $JVJ$, so
\begin{equation}
\fer{a40}= -P_e JVJ \left[ \e^{-\i t(L_0-e)} + \e^{\i t(L_0-e)}\right] VP_e.
\label{a41}
\end{equation}
This is an even function in $t$ so it does not contribute to the integral in \fer{a37}. It follows that
\begin{equation}
2\RE\Lambda_e(\epsilon) =\i \int_\r\d t\ \e^{-\epsilon |t|}\sgn(t) P_e\left[ V\e^{-\i t(L_0-e)} V + JV\e^{-\i t(L_0+e)}VJ\right] P_e.
\label{a42}
\end{equation}
In order to calculate matrix elements of \fer{a42} we use the following fact. If $M$ is any operator affiliated with $\mm$ and if $B\in\b(\k)$ then $\bbbone_\k\otimes B\otimes\bbbone_{\h_{\rm S}}$ leaves the domain of $M$ invariant, and the two operators commute strongly on that domain. (This holds since $\bbbone_\k\otimes B\otimes\bbbone_{\h_{\rm S}}\in\mm'$.) A similar statement holds for $M'$ affiliated with $\mm'$ and $B\otimes\bbbone_\k\otimes\bbbone_{\h_{\rm S}}$.

Let $(k,l)$, $(k',l')$ be indices s.t. $E_{k,l}=E_{k',l'}=e$. Then
\begin{eqnarray}
\lefteqn{\scalprod{\varphi_k\otimes\varphi_l\otimes\Phi_\beta}{\ 2\RE\Lambda_e(\epsilon) \ \varphi_{k'}\otimes\varphi_{l'}\otimes\Phi_\beta}} \nonumber \\
&=&\i \int_\r\d t\ \e^{-\epsilon |t|}\sgn(t)\Big[ \delta_{l,l'}\ \scalprod{\varphi_k\otimes\varphi_l\otimes\Phi_\beta}{V \e^{-\i t(L_0-e)} V\varphi_{k'}\otimes\varphi_{l}\otimes\Phi_\beta}\nonumber\\
&&+ \delta_{k,k'}\ \scalprod{\varphi_{l'}\otimes\varphi_k\otimes\Phi_\beta}{V \e^{\i t(L_0+e)} V\varphi_{l}\otimes\varphi_k \otimes\Phi_\beta}\Big].
\label{a43}
\end{eqnarray}
Consider the case $e=0$. Since all eigenvalues of $H$ are simple we must have $k=l$ and $k'=l'$. The Kronecker deltas $\delta_{l,l'}$ and $\delta_{k,k'}$ in \fer{a43} force all indices to be the same: $k=l=k'=l'$. Thus the integrand in \fer{a43} becomes an odd function of $t$ and the value of the integral is zero.

This shows that $\RE\Lambda_0(\epsilon)=0$, so $\Lambda_0(\epsilon)= \i\Gamma_0(\epsilon)$, where
\begin{equation}
\Gamma_0(\epsilon):=\frac{1}{2i}[\Lambda_0(\epsilon)-\Lambda_0(\epsilon)^*] = P_0I\Pbar_0 \frac{\epsilon}{(\overline{L}_0)^2+\epsilon^2}\Pbar_0IP_0 \geq 0
\label{a45}
\end{equation}
is manifestly a positive definite operator. (We write $\overline{L}_0$ for $\overline{L}_0^0$ in \fer{a45}.)

Next we verify that $\Gamma_0(\epsilon)$ has real matrix elements in the basis $\{\varphi_i\otimes\varphi_i\otimes\Phi_\beta\}$. We use again the properties of $J$, as in the proof of (a), to see that $J\Gamma_0(\epsilon)J=\Gamma_0(\epsilon)$ (this can also be viewed as a consequence of assertion (a) in the theorem and the fact that $\Lambda_0(\epsilon)=\i\Gamma_0(\epsilon)$). It follows that
\begin{eqnarray*}
\lefteqn{
\scalprod{\varphi_i\otimes\varphi_i\otimes\Phi_\beta}{\Gamma_0(\epsilon)  \varphi_j\otimes\varphi_j\otimes\Phi_\beta}}\\
&=& \overline{\scalprod{J\varphi_i\otimes\varphi_i\otimes\Phi_\beta}{\Gamma_0(\epsilon)J \varphi_j\otimes\varphi_j\otimes\Phi_\beta}}\\
&=&\overline{\scalprod{\varphi_i\otimes\varphi_i\otimes\Phi_\beta}{ \Gamma_0(\epsilon) \varphi_j\otimes\varphi_j\otimes\Phi_\beta}},
\end{eqnarray*}
so the matrix elements are real.

We have now shown assertion (b) of Theorem \ref{thm3}, modulo the easy proof of Lemma \ref{lemma1}.

{\bf Proof of Lemma \ref{lemma1}.\ } First note that $\e^{\i tL_0}M\e^{-\i tL_0} P_e$, $P_e \e^{\i tL_0}M\e^{-\i tL_0}$ and $P_e JMJ$ are bounded. Let $\chi_n=\chi(|M|\leq n)$ be a spectral cutoff operator. Then $M_n:=\chi_n M\in\mm$, and $J M_n J\in\mm'$, for all $n$. The l.h.s. of \fer{a38} is we weak limit of
\begin{equation}
P_e\e^{\i tL_0}M_n \e^{-\i tL_0} J M_n JP_e = P_e J M_n J \e^{\i tL_0} M_n \e^{-\i tL_0}  P_e,
\label{a39}
\end{equation}
as $n\rightarrow\infty$. The equality holds since $\e^{\i tL_0}\ \cdot\ \e^{-\i tL_0}$ leaves $\mm$ invariant. The r.h.s. of \fer{a39} converges weakly to the r.h.s. of \fer{a38}, as $n\rightarrow\infty$. \hfill $\blacksquare$

\subsection{Proof of Theorems \ref{thm2} and \ref{thm5}}

We prove Theorem \ref{thm1} below, which covers the proofs of both Theorems \ref{thm2} and \ref{thm5}.

Let $L_0$ be a selfadjoint operator on a Hilbert space ${\cal H}$ and let $W$ be an operator on ${\cal H}$ s.t. $K_\lambda=L_0+\lambda W$ defines a closed operator for $\lambda\in U$, where $U\subset \r$ is a neighbourhood of the origin.
Assume that we have $K_\lambda\psi_\lambda = e\psi_\lambda$, for some $e\in\r$ and for all $\lambda\in U$, where $\psi_\lambda\in\h$. Let $P$ be the selfadjoint projection onto the eigenspace of $L_0$ associated with $e$, and set $\Pbar=\bbbone-P$. We allow the case $\dim P=\infty$. Let $\Lbar_0=\Pbar L_0\Pbar\upharpoonright_{\ran \Pbar}$ and $\Kbarlambda=\Pbar K_\lambda \Pbar\upharpoonright_{\ran \Pbar}$ denote the restrictions of $L_0$ and $K_\lambda$ to $\ran \Pbar$, respectively.  We make the following hypotheses:
\begin{itemize}
\item[{\bf H0}] There is a sequence $\zeta_n$ in the open upper complex half plane $\cx_+$ converging to a $\zeta\in\cx_+$, such that all $\zeta_n$ and $\zeta$ belong to the resolvent sets of $\Kbarlambda$ for all $\lambda\in U$.
\item[{\bf H1}] $PW$ and $WP$ are bounded operators on $\h$.
\item[{\bf H2}] $\lambda\mapsto PW\Pbar(\Kbarlambda-\zeta_n)^{-1}\Pbar$ is continuous at $\lambda=0$ as a map from $\rx$ to the bounded operators on $\h$, for every $\zeta_n$.
\item[{\bf H3}] $\lambda\mapsto \psi_\lambda$ is differentiable at $\lambda=0$ as a map from $\rx$ to $\h$, with derivative at zero denoted by $\psi'_0$.
\end{itemize}

\begin{theorem}
\label{thm1}
Assume hypotheses H0--H3. Then we have
\begin{equation}
\lim_{\epsilon\downarrow 0} PW\Pbar({\overline L}_0-e-\i\epsilon)^{-1}\Pbar WP\psi_0=PWP\psi'_0.
\label{2}
\end{equation}
If H0 hods for a sequence $\zeta_n\in\cx_-$ converging to a $\zeta\in\cx_-$ then \fer{2} holds with the limit replaced by $\lim_{\epsilon\uparrow 0}$. In case H0 holds for two sequences, one in the upper half plane, the other in the lower one, then \fer{2} holds with the limit replaced by $\lim_{\epsilon\rightarrow 0}$.
\end{theorem}
{\bf Remark.\ } One can show that under suitable regularity conditions on $W$, the operator $PW\Pbar(\overline{L}_0-e-\i\epsilon)^{-1}\Pbar WP$ has limits as $\epsilon\downarrow 0$ and as $\epsilon\uparrow 0$. Theorem \ref{thm1} does not presuppose the existence of these limiting operators, though.

The {\it proof of Theorem \ref{thm2}} is an application of Theorem \ref{thm1}, where $K_\lambda$ is the selfadjoint standard Liouville operator $L_\lambda$ \fer{a4} and where $e=0$ and $\psi_\lambda =\Omega_{\beta,\lambda}$ is the interacting KMS state \fer{a20}. H0 is satisfied since $L_\lambda$ is selfadjoint and H1-H3 follow from A1$_0$-A3. To prove Theorem \ref{thm5} we apply Theorem \ref{thm1} with $K_\lambda$ given in Condition C after \fer{n1}. Since $W$ one easily sees that conditions H0-H2 are met, and H3 is satisfied by the assumption in Theorem \ref{thm5}.

{\it Proof of Theorem \ref{thm1}.\ }
In the spirit of a {\it Feshbach-type argument} we project the equation $(K_\lambda-e) \psi_\lambda=0$ onto $\ran P$ and $\ran \Pbar$,
\begin{eqnarray}
PWP\psi_\lambda &=& - P W \Pbar\psi_\lambda, \label{3}\\
\Pbar(K_\lambda-e)\Pbar\psi_\lambda&=& -\lambda \Pbar W P\psi_\lambda \label{4}.
\end{eqnarray}
Take a fixed element $\zeta_n$ from the sequence in H0. We add the vector $(e-\zeta_n)\Pbar \psi_\lambda$ to both sides of \fer{4} and ``solve'' for $\Pbar\psi_\lambda$,
\begin{equation}
\Pbar\psi_\lambda = -\lambda \Pbar(\Kbarlambda-\zeta_n)^{-1}\Pbar W P \psi_\lambda +(e-\zeta_n) \Pbar(\Kbarlambda-\zeta_n)^{-1}\Pbar\psi_\lambda.
\label{5}
\end{equation}
Substitution in equation \fer{3} results in
\begin{equation}
 PWP\psi_\lambda = \lambda PW \Pbar(\Kbarlambda-\zeta_n)^{-1}\Pbar WP\psi_\lambda -(e-\zeta_n) PW\Pbar (\Kbarlambda-\zeta_n )^{-1}\Pbar\psi_\lambda.
\label{6}
\end{equation}
Since $\Pbar\psi_0=0$ equation \fer{6} can be written as 
\begin{equation}
-PW\Pbar \frac{\psi_\lambda-\psi_0}{\lambda}=PW\Pbar(\Kbarlambda-\zeta_n)^{-1}\Pbar WP\psi_\lambda -(e-\zeta_n) PW\Pbar(\Kbarlambda-\zeta_n)^{-1}\Pbar\frac{\psi_\lambda-\psi_0}{\lambda}.
\label{8}
\end{equation}
The l.h.s. of \fer{8} equals $PWP\frac{\psi_\lambda-\psi_0}{\lambda}$ because we have $PW\psi_\lambda=0$ for all $\lambda$ (see \fer{3}). In the limit $\lambda\rightarrow 0$ equation \fer{8} is
\begin{equation}
PWP\psi'_0 = PW\Pbar (\overline{L}_0-\zeta_n)^{-1}\Pbar WP\psi_0 -(e-\zeta_n) PW\Pbar (\overline{L}_0-\zeta_n)^{-1} \Pbar\psi'_0.
\label{9}
\end{equation}
The map $z\mapsto PW\Pbar (\overline{L}_0-z)^{-1}\Pbar WP\psi_0 -(e-z) PW\Pbar (\overline{L}_0-z)^{-1} \Pbar\psi'_0$ is analytic in $z\in\cx_+$ and, by \fer{9}, it is constant on the convergent sequence $\zeta_n$. We conclude that equality \fer{9} holds with $\zeta_n$ replaced by any $z\in\cx_+$. Choose $z=e+\i\epsilon$, with $\epsilon>0$,
\begin{equation}
PWP\psi'_0 = PW\Pbar (\overline{L}_0-e-\i\epsilon)^{-1}\Pbar WP\psi_0 +\i\epsilon PW\Pbar (\overline{L}_0-e-\i\epsilon)^{-1} \Pbar\psi'_0.
\label{9.1}
\end{equation}
We now take $\epsilon\downarrow 0$. Relation \fer{2} follows since $\i\epsilon\Pbar (\overline{L}_0-e-\i\epsilon)^{-1}\Pbar$ converges strongly to zero (irrespective of the sign of $\epsilon$).

In case H0 holds for a sequence $\zeta_n$ belonging to $\cx_-$ the same argument shows \fer{2} with the limit replaced by $\lim_{\epsilon\uparrow 0}$. 
\hfill $\blacksquare$

\subsection{Proof of Theorem \ref{thm4}}
It is shown in \cite{DJP} that $\Omega_{\beta,0}\in\dom(\e^{-\beta\lambda V/2})$ implies Condition B2. All other conditions are very easily seen to hold. The analytic extension of the resolvent is
\begin{equation}
\Pbar_0 (\Lbarlambda-\i\epsilon)^{-1}\Pbar_0 =\Pbar_0 (\overline{L}_0-\i\epsilon)^{-1}\Pbar_0 \sum_{n=0}^\infty (-\lambda)^n \left[\, \overline{I} (\overline{L}_0-\i\epsilon)^{-1}\right]^n\Pbar_0,
\label{a55}
\end{equation}
provided $|\lambda|<\epsilon/\|\overline{I}\|$. The analytic extension of the perturbed KMS state is given (modulo normalization) by
\begin{equation}
\Omega_{\beta,\lambda} = \sum_{n=0}^\infty (-\lambda)^n\int{\cdots}\int_{T_{\beta,n}} \!\! \d\beta_1\cdots\d\beta_n\ \e^{-\beta_1 L_0} V\cdots \e^{-\beta_n L_0} V\Omega_{\beta,0},
\label{a56}
\end{equation}
where $T_{\beta,n}=\{(\beta_1,\ldots,\beta_n)\in\r^n\ | \ \beta_i\geq 0, \beta_1+\cdots+\beta_n\leq\beta/2\}$. Expansion \fer{a56} is due to Araki, \cite{Araki}, see also \cite{DJP,BR}. \hfill $\blacksquare$

\subsection{Proof of Theorem \ref{thmbosons1}}
\label{sectproofthmbosons1}

A verification of conditions B1, B2, A1--A3 is quite standard. We outline the main steps.
One proves \cite{RSII} (Theorem X.44) that $V$ and $JVJ$ are well defined symmetric operators with a core for essential selfadjointness given by $\k\otimes\k\otimes\ff_0\otimes\ff_0$, where $\ff_0$ is the set of all finite linear combinations of functions in $\otimes_{\rm sym}^n C_0^\infty(\r^3,\d^3k)$, with variable $n$ (``the finite particle space over $C_0^\infty$ test functions'').
An argument as given e.g. in \cite{BD} (Theorem 3.13) shows that $L_0+\lambda V$ and $L_0+\lambda V-\lambda JVJ$ are essentially selfadjoint on $\k\otimes\k\otimes\ff_0\otimes\ff_0$. Thus, Condition B1 holds.

Condition B2 follows from Condition A3 which we show to hold below.

To verify Condition A1$_e$ we note that $P_e=\chi_{L_{\rm S}=e} \otimes P_{\rm R}$, where $\chi_{L_{\rm S}=e}$ is the spectral projection onto the eigenspace corresponding to the eigenvalue $e$ of $L_{\rm S}$, and $P_{\rm R}=|\Phi_\beta\rangle\langle\Phi_\beta|$ is the orthogonal projection onto $\cx \Phi_\beta$. Denote by $N_\beta=N\otimes\bbbone_{\ff_+}+\bbbone_{\ff_+}\otimes N$, where $N=\int_{\r^3}a^*(k)a(k)\d^3k$ is the number operator on $\ff_+$. Since $N_\beta \Phi_\beta=0$ the standard bound
\begin{equation}
\| I (N_\beta+1)^{-1}\|=C<\infty
\label{a75}
\end{equation}
shows that A1$_e$ holds for all eigenvalues $e$ of $L_{\rm S}$.

To verify Condition A2 we use the resolvent identity
\begin{eqnarray}
\lefteqn{
P_0I\Pbar_0(\Lbar_\lambda-\i\epsilon)^{-1}\Pbar_0 }\nonumber\\
&=& P_0I\Pbar_0(\Lbar_0-\i\epsilon)^{-1}\Pbar_0 + \lambda P_0I\Pbar_0(\Lbar_0-\i\epsilon)^{-1} \overline{I} (\Lbar_\lambda-\i\epsilon)^{-1}\Pbar_0 .
\label{a76}
\end{eqnarray}
One easily verifies that for any $\nu\geq 0$,
\begin{equation}
I P(N_\beta\leq\nu)\in {\rm Ran\,}P(N_\beta\leq \nu+2),
\label{a73}
\end{equation}
where $P(N_\beta\leq\nu)$ is the spectral projection of $N_\beta$ corresponding to the interval $[0,\nu]$. It follows that the norm of the second term on the r.h.s. of \fer{a76} is bounded above by $C |\lambda|\epsilon^{-2}\|P(N_\beta\leq 2)I\|$ (we commute $P(N_\beta\leq 2)$ through the resolvent $(\Lbar_0-\i\epsilon)^{-1}$). This shows that A2 holds.

Finally we check Condition A3. A Dyson series expansion allows us to write
\begin{eqnarray}
\lefteqn{
\e^{-\beta(L_0+\lambda V)/2}\e^{\beta L_0/2}\Omega_{\beta,0}}\nonumber\\
&=&\sum_{n=0}^\infty \Big(- \frac{\lambda}{2} \Big)^n\int_{0\leq s_n\leq\cdots\leq s_1\leq \beta/2} \d s_1\cdots \d s_n \ \alpha_0^{\i s_n}(V)\cdots \alpha_0^{\i s_1}(V)\Omega_{\beta,0},\ \ \ \ \ \
\label{a77}
\end{eqnarray}
where $\Omega_{\beta,0}$ is the uncoupled KMS state, \fer{a3.9}, and where, for $z\in\cx$,
\begin{equation}
\alpha_0^z(V)=\e^{\i z L_0} V \e^{-\i zL_0}.
\label{a78}
\end{equation}
Formula \fer{a77} holds if the series on the r.h.s. converges, \cite{Matthias}, Appendix B.1. The relation
\begin{equation}
\e^{\i zL_0} a_\beta^*(k)\e^{-\i zL_0} = \e^{\i z|k|} a_\beta^*(k)
\label{a79}
\end{equation}
together with its adjoint allows us to estimate the integrand of \fer{a77} by
\begin{eqnarray}
\lefteqn{
\| \alpha_0^{\i s_n}(V)\alpha_0^{\i s_{n-1}}(V)\cdots \alpha_0^{\i s_1}(V)\Omega_{\beta,0}\| }\nonumber\\
&=&\|\alpha_0^{\i s_n}(V)P(N_\beta\leq 2n)\alpha_0^{\i s_{n-1}}(V)P(N_\beta\leq 2(n-1)) \cdots \alpha_0^{\i s_1}(V)\Omega_{\beta,0}\| \nonumber\\
&\leq& C(s_1)\cdots C(s_n) 2^n n!,
\label{a80}
\end{eqnarray}
where $C(s)$ is defined in \fer{a64}. By \fer{a70} we have $\fer{a80}\leq (2C)^n n!$, so the series \fer{a77} converges provided
\begin{equation}
 C\beta|\lambda|/2<1.
\label{a81}
\end{equation}
For $\lambda$ satisfying the bound \fer{a81} the map $\lambda\mapsto\Omega_{\beta,\lambda}$ has an analytic extension, given by \fer{a77}.

{\bf Remark.\ } If $h_2=0$ then $V$ is relatively bounded w.r.t. $N_\beta^{1/2}$ and the $n!$ in the r.h.s. of \fer{a80} can be replaced by $(n!)^{1/2}$. Thus the series \fer{a77} converges for all values of $\lambda\in\rx$.
\hfill $\blacksquare$

\subsection{Proof of Theorem \ref{thmaccobs}}
\label{sectproofthmaccobs}

It is useful to pass from $L^2(\r^3,\d^3x)$ to $L^2(\r^3,\d\mu\d \uk^\perp)$, where $\uk^\perp=(k^2,k^3)$ and $\mu\in\rx$, according to the isometric isomorphism
\begin{equation}
f \mapsto (2\pi)^{-1/2}\int_\rx\d\kappa \e^{-\i\mu\kappa}\widehat f (\omega_\perp\sinh\kappa,\uk^\perp),
\label{b22}
\end{equation}
where $\widehat f$ is the Fourier transform of $f$ and $\omega_\perp=(\uk^\perp+m^2)^{1/2}$. In the new space, the operator \fer{b8.1} becomes $L_{\rm R}=\d\Gamma(\mu)$, and the interaction \fer{b21} becomes
\begin{equation}
V= G\otimes\bbbone_\k\otimes  \phi(\sigma),
\label{b23}
\end{equation}
where $\phi$ is the usual zero temperature bosonic field operator acting on the bosonic Fock space over $L^2(\r^3,\d\mu\d\uk^\perp)$, smeared out with
\begin{equation}
\sigma(\mu,\uk^\perp) = (2\pi)^{1/2}\int_\rx\d\kappa\e^{-\i\mu\kappa}\widehat\rho(\omega_\perp\sinh\kappa,\uk^\perp).
\label{b24}
\end{equation}
One can use a standard Nelson-Commutator-Theorem argument to show that Condition B1 is satisfied (for all values of $\lambda$).

To check Conditions B2 and A3 we use the Dyson series \fer{a77}, as for the thermal Bosons. The Dyson series converges due to the assumption $\sup_{0\leq s\leq\pi}\|\e^{-s K}G\e^{s K}\|<\infty$ and the fact that $\e^{-\pi\mu}\sigma\in L^2(\r^3,\d\mu\d\uk^\perp)$. The latter fact follows, via the transforms \fer{b24} and \fer{b22}, from the Bisognano--Wichmann theorem, which asserts that $\varphi[g]\Omega_{\rm R}$ is in the domain of the operator $\e^{-\pi L_{\rm R}}$, since $g(x)=\delta(x^0)\rho(\ux)$ is supported in ${\cal W}_{\rm R}$. The convergent Dyson series defines the perturbed KMS vector which is entire analytic in $\lambda$. In particular, Conditions B2 and A3 are satisfied.

Conditions A1$_e$ and A2 are verified just as for thermal Bosons, Section \ref{sectproofthmbosons1}. \hfill $\blacksquare$

\subsection{Proof of Theorem \ref{thmdegspec}}
\label{proofthmdegspec}

Since $L_{\rm S}=0$ we have $P=\bbbone_{\cx^2}\otimes\bbbone_{\cx^2}\otimes P_{\Omega}\otimes P_{\Omega}$, where $\Omega$ is the vacuum vector in $\ff_+$. By using the definition \fer{a8}, the explicit form of $V$, \fer{d1}, and the Araki--Woods representation \fer{a62.1} we arrive at
\begin{eqnarray}
\lefteqn{\Lambda_0(\epsilon)=\frac{\i}{2} P\Big[ G\otimes\bbbone_{\cx^2}\otimes \big\{ a(\sqrt{1+\rho}\ g)\otimes\bbbone_{\ff_+} +\bbbone_{\ff_+}\otimes a(\sqrt{\rho}\, \overline g)\big\}}\nonumber\\
&&\ \ \ -\bbbone_{\cx^2}\otimes\cc G\cc\otimes\big\{ \bbbone_{\ff_+}\otimes a(\sqrt{1+\rho}\ \overline g)+ a(\sqrt{\rho}\ g)\otimes\bbbone_{\ff_+}\big\}\Big]\nonumber\\
&&\times\int_0^\infty \d t \ \e^{-\epsilon t} \e^{-\i t L_{\rm R}}
\Big[ G\otimes\bbbone_{\cx^2}\otimes \big\{ a^*(\sqrt{1+\rho}\ g)\otimes\bbbone_{\ff_+} +\bbbone_{\ff_+}\otimes a^*(\sqrt{\rho}\, \overline g)\big\}\nonumber\\
&&\ \ \ -\bbbone_{\cx^2}\otimes\cc G\cc\otimes\big\{ \bbbone_{\ff_+}\otimes a^*(\sqrt{1+\rho}\ \overline g)+ a^*(\sqrt{\rho}\ g)\otimes\bbbone_{\ff_+}\big\}\Big]P.
\label{d6}
\end{eqnarray}
We have represented the reslovent $(\overline L_0-\i\epsilon)^{-1}$ in integral form. Taking into account $\e^{-\i t L_{\rm R}}(a^*(f)\otimes \bbbone_{\ff_+}) \e^{\i tL_{\rm R}} = a^*(\e^{-\i t\omega} f)\otimes \bbbone_{\ff_+}$ and $\e^{-\i t L_{\rm R}}(\bbbone_{\ff_+}\otimes a^*(f)) \e^{\i tL_{\rm R}} = \bbbone_{\ff_+}\otimes a^*(\e^{\i t\omega} f)$, and the formula $\i\int_0^\infty \d t\ \e^{-\epsilon t}\e^{\pm \i\omega t}=-(\pm\omega+\i\epsilon)^{-1}$, expression \fer{d6} reduces to (contractions!)
\begin{eqnarray}
2\Lambda_0(\epsilon) &=& G^2\otimes\bbbone_{\cx^2}\scalprod{g}{\Big[\frac{1}{\omega-\i\epsilon} +\rho \frac{2\i\epsilon}{\omega^2+\epsilon^2}\Big ]g}\nonumber\\
&& + \bbbone_{\cx^2}\otimes \cc G^2\cc \scalprod{g}{\Big[\frac{1}{-\omega-\i\epsilon} +\rho \frac{2\i\epsilon}{\omega^2+ \epsilon^2}\Big ]g}\nonumber\\
&&-2 G\otimes \cc G \cc \scalprod{g}{\Big[ \sqrt{\rho(1+\rho)} \frac{2\i\epsilon}{\omega^2+\epsilon^2}\Big ]g},
\label{d7}
\end{eqnarray}
viewed as an operator on $\cx^2\otimes\cx^2$. Next we use the formula
\begin{equation}
\int_0^\infty \d r\ h(r)\frac{2\i\epsilon}{r^2+\epsilon^2} \longrightarrow \i\pi h(0),
\label{d8}
\end{equation}
in the limit $\epsilon\downarrow 0$, for any bounded and continuous function $h$ on $\r_+$. In order to calculate the limit $\epsilon\downarrow 0$ of the second term of the first summand in the r.h.s. of \fer{d7} we use \fer{d8} with
\begin{equation}
h(|k|) = -\frac{|k|^2}{\e^{\beta|k|}-1} |k|^{2p}\int_{S^2}\d\sigma\ \frac{|g(|k|,\sigma)|^2}{|k|^{2p}},
\label{d9}
\end{equation}
where use spherical coordinates in $\r^3$, and the factor $|k|^2$ is the Jacobian. For $p\geq -1/2$ the function $h$ is bounded and continuous. One proceeds similarly for the other two summands in \fer{d7}. This shows parts 2 and 3 of Theorem \ref{thmdegspec}.

If $-1<p<-1/2$ then it is readily seen that $\int_0^\infty \d r\ h(|k|) \frac{2\i\epsilon}{|k|^2+\epsilon^2}\rightarrow\i \infty$, as $\epsilon\downarrow 0$, and where $h$ is given by \fer{d9}. This shows assertion 1 of Theorem \ref{thmdegspec}. \hfill $\blacksquare$

\subsection{Proofs of Theorem \ref{degthm2} and of Proposition \ref{degprop}}

{\it Proof of Theorem \ref{degthm2}}.\
We introduce the following ordered orthonormal basis for $\ker L_{\rm S}$:
\begin{equation}
\big\{ \varphi_1\otimes\varphi_1, \varphi_2\otimes\varphi_2, \varphi_3\otimes\varphi_3, \varphi_2\otimes\varphi_3, \varphi_3\otimes\varphi_2 \big\},
\label{onb}
\end{equation}
where the $\varphi_j$ constitute the canonical basis of $\cx^3$. A somewhat longish calculation (that is carried out as in the previous section) yields the following expression for $\Lambda_0$ in the basis \fer{onb}
\begin{equation}
\Lambda_0 =\frac{\i s}{2\sinh(\beta\Delta/2)}
\left[
\begin{array}{ccccc}
\frac{a^2+b^2}{\e^{\beta\Delta/2}}   & -a^2  & -b^2  & -ab  &  -ab \smallskip\\
-a^2  & a^2 \e^{\beta\Delta/2} & 0 & \frac{ab}{2}\zeta & \frac{ab}{2}\overline\zeta\smallskip\\
-b^2  & 0 & b^2\e^{\beta\Delta/2}  & \frac{ab}{2}\overline\zeta & \frac{ab}{2}\zeta \smallskip\\
-ab & \frac{ab}{2}\zeta & \frac{ab}{2}\overline\zeta & \frac{a^2\overline\zeta +b^2\zeta}{2} & 0\smallskip\\
-ab & \frac{ab}{2}\overline\zeta & \frac{ab}{2}\zeta & 0 & \frac{a^2\zeta +b^2\overline\zeta}{2}
\end{array}
\right],
\label{d11}
\end{equation}
where $s=\pi \Delta^2 \int_{S^2}\d\sigma |g(\Delta,\sigma)|^2$ and
$\zeta = \e^{\beta \Delta/2}-\frac{4\i}{s}\sinh(\beta\Delta/2) \
{\rm P.V.}\scalprod{g}{\frac{1+\rho_\beta}{\Delta-\omega}g}$,
$\omega(k)=|k|$ and $\rho_\beta$ is defined after \fer{a61}. The
kernel of $\Lambda_0$ is spanned by the two vectors
\begin{equation}
\Psi =
\left[
\begin{array}{c}
\e^{\beta\Delta/2}\\
1\\
1\\
0\\
0
\end{array}
\right]
\mbox{\ \ \ and\ \ \ }
\chi = \frac{a}{b}
\left[
\begin{array}{c}
\e^{\beta\Delta/2}\\
1-(b/a)^2\\
0\\
1\\
1
\end{array}
\right]
\label{d12}
\end{equation}
(of course, $\Psi$ is proportional to the Gibbs state $\Psi_\beta$
of the small system, \fer{a12}).

In case $g(k)$ satisfies \fer{d2} with $p=-1/2$ the matrix
$\Lambda_0$ is given by the following modification of \fer{d11}.
Add to the r.h.s. of \fer{d11} the diagonal matrix $2i\pi
c^2\delta$\,diag$(0,1,1,1,1)$, where
$\delta=\beta^{-1}\lim_{r\rightarrow 0} r \int_{S^2}\d\sigma\,
|g(r,\sigma)|^2$ and replace the zeroes in the matrix \fer{d11} by
$-2\i\pi c^2\delta\ \frac{2\sinh(\beta\Delta/2)}{\i s}$.

\smallskip
{\it Proof of Proposition \ref{degprop}}.\ We write $P,\Pbar$
instead of $P_0,\Pbar_0$ in this proof. It follows from \fer{d19}
and \fer{d18.9} that
\begin{equation}
\Lambda_0'= \lim_{\epsilon\downarrow 0} PI\Pbar
(\Lbar_0-\i\epsilon)^{-1}\Pbar
I\Pbar(\Lbar_0-\i\epsilon)^{-1}\Pbar I\Pbar
(\Lbar_0-\i\epsilon)^{-1}\Pbar IP. \label{d30}
\end{equation}
(For a translation analytic system this limit is easily seen to
exist, compare with \fer{d18.9}). For $b=0$ expression \fer{d11}
reduces considerably and one checks that the kernel of $\Lambda_0$ is spanned
by the two vectors $\Psi$ given in \fer{d12} and $\Psi_0=[0\ 0\ 1\ 0\ 0]^{\rm
t}\cong\varphi_3\otimes\varphi_3$. We apply analytic perturbation
theory (in $\lambda^2$) to the matrix
$\Lambda_0+\lambda^2\Lambda_0'$. The correction of order
$\lambda^2$ to the doubly degenerate eigenvalue zero of
$\Lambda_0$ is given by the eigenvalues of the matrix
\begin{equation}
D=\lambda^2\left[
\begin{array}{cc}
\scalprod{\Psi}{\Lambda_0'\Psi} & \scalprod{\Psi}{\Lambda_0'\Psi_0} \\
\scalprod{\Psi_0}{\Lambda_0'\Psi} & \scalprod{\Psi_0}{\Lambda_0'\Psi_0}
\end{array}
\right].
\label{d31}
\end{equation}
We have $F(L_\lambda(\tau))P\e^{-\beta(L_0+\lambda
V)/2}\Omega_{\beta,0}=0$ for all values of $\lambda$, where
$\Omega_{\beta,0}$ is the non-interacting KMS vector, c.f.
\fer{a20}. An expansion of this relation yields for the order in
$\lambda^2$ the equation
\begin{equation}
\Lambda_0\chi_1 + \Lambda_0'\Psi =0,
\label{d31.1}
\end{equation}
where
\begin{equation*}
\chi_1 = \frac 14\int_0^\beta\d\beta_1\int_0^{\beta_1}\d\beta_2
PV\e^{(\beta_2-\beta_1)L_0/2} V\Omega_{\beta,0}.
\end{equation*}
By taking the inner product of $\Psi$ and $\Psi_0$ with \fer{d31.1}
we conclude that
$\scalprod{\Psi}{\Lambda_0'\Psi}=\scalprod{\Psi_0}{\Lambda_0'\Psi}=0$.
Thus, since $\det D=0$ and $\tr D=
\lambda^2\scalprod{\Psi_0}{\Lambda_0'\Psi_0}$ the spectrum of $D$
is $\{0, \lambda^2 \scalprod{\Psi_0}{\Lambda_0'\Psi_0}\}$. In
order to calculate the matrix element
$\scalprod{\Psi_0}{\Lambda_0'\Psi_0}$ we represent the resolvents
in \fer{d30} by integrals,
\begin{equation}
\Lambda_0'= -\i \lim_{\epsilon\downarrow 0}\int_0^\infty\!\! \d
t_1\int_0^\infty\!\! \d t_2\int_0^\infty\!\!\d t_3 \
\e^{-\epsilon(t_1+t_2+t_3)}  PI\Pbar I(s_1)\Pbar I(s_2)\Pbar
I(s_3)P, \label{d32}
\end{equation}
where $I(t)=\e^{\i tL_0}I\e^{-\i tL_0}$, $s_1=-t_1$,
$s_2=-t_1-t_2$, $s_3=-t_1-t_2-t_3$. Since $PIP=0$ we may drop the
projections $\Pbar$ in \fer{d32} except for the middle one. An
explicit calculation gives
\begin{eqnarray}
\lefteqn{
I(s_2)I(s_3)\Psi_0\otimes\Phi_\beta}\nonumber\\
 &=& c
\begin{bmatrix} a\e^{-\i s_2\Delta}\\ 0\\ c
\end{bmatrix}
\otimes\varphi_3\otimes T_1
-c^2\varphi_2\otimes\varphi_2\otimes T_2
+c\varphi_3\otimes
\begin{bmatrix} a\e^{\i s_2\Delta} \\ 0 \\ c
\end{bmatrix}
\otimes T_3,
\label{d33}
\end{eqnarray}
where $T_1, T_2, T_3$ are vectors in the Hilbert space of the
reservoir. Similarly,
\begin{eqnarray}
\lefteqn{
I(s_1)I\Psi_0\otimes\Phi_\beta}\nonumber\\
 &=& c
\begin{bmatrix} a\e^{-\i s_1\Delta}\\ 0\\ c
\end{bmatrix}
\otimes\varphi_3\otimes S_1
-c^2\varphi_2\otimes\varphi_2\otimes S_2
+c\varphi_3\otimes
\begin{bmatrix} a\e^{\i s_1\Delta} \\ 0 \\ c
\end{bmatrix}
\otimes S_3,
\label{d34}
\end{eqnarray}
where the $S_j$ are obtained from the $T_j$ by replacing $s_2$ by
$s_1$ and setting $s_3=0$. One checks that
\begin{equation*}
\lim_{\epsilon \rightarrow 0}\int_0^\infty \d t_3\ \e^{-\epsilon
t_3} \scalprod{\Psi_0\otimes\Phi_\beta}{I\ I(s_1) P
I(s_2)I(s_3)\Psi_0\otimes\Phi_\beta}=0,
\end{equation*}
so the middle $\Pbar$ in \fer{d32} can also be dropped (we may
take the epsilons multiplying $t_1, t_2, t_3$ in \fer{d32}
individually to zero). The inner product
$\scalprod{\fer{d34}}{\fer{d33}}$ equals
\begin{equation}
a^2c^2 \left(\e^{\i t_2\Delta}\scalprod{S_1}{T_1} +\e^{-\i
t_2\Delta}\scalprod{S_3}{T_3}\right) +c^4
\left(\scalprod{S_1+S_3}{T_1+T_3} +\scalprod{S_2}{T_2}\right).
\label{d35}
\end{equation}
A pretty lengthy calculation shows that the term in \fer{d35}
which is proportional to $a^2c^2$ yields the following
contribution to $\scalprod{\Psi_0}{\Lambda'_0\Psi_0}$,
\begin{eqnarray}
\lefteqn{\i a^2c^2 \xi_1(\Delta):=}\label{xi1}\\
 &&2\i\pi
a^2c^2\Delta^2\int_{S^2}\d\sigma\!\!\int_{S^2} \d\sigma'\!\!
\int_\Delta^\infty \!\!\d r |g(\sigma,r)|^2
|g(\sigma',r-\Delta)|^2 \rho_\beta(r-\Delta)(1+\rho_\beta(r)),
\nonumber
\end{eqnarray}
where $\rho_\beta(r)=(\e^{\beta r}-1)^{-1}$. While this
contribution is purely imaginary the one coming from the term in
\fer{d35} proportional to $c^4$ is not; its imaginary part is
obtained again by a quite longish calculation and it is given by
\begin{eqnarray}
\lefteqn{c^4 \xi_2:=}\label{xi2}\\
 &&\pi c^4 \int_{S^2}\d\sigma\!\!\int_{S^2} \d\sigma'\!\!
\int_0^\infty \!\!\d r\  r^2 |g(\sigma,r)|^2 |g(\sigma',r)|^2
\Big( 2\sqrt{ \rho_\beta(r)(1+\rho_\beta(r))} -1-2\rho_\beta(r)
\Big)^2.
 \nonumber
\end{eqnarray}
This concludes the proof of Proposition \ref{degprop}. \hfill
$\blacksquare$

\quad

{\bf Acknowledgement} 
I thank S. DeBi\`evre for stimulating and encouraging discussions, and J. Fr\"ohlich and I.M. Sigal for all they have taught me.


\begin{thebibliography}{99}
\bibitem{A}
Abou-Salem, W.K.: {\it On the quasi-static evolution of non-equilibrium steady states}, preprint, mp-arc 05-341

\bibitem{AF}
Abou-Salem, W.K., Fr\"ohlich, J.: {\it Adiabatic theorems and reversible isothermal processes}  Lett. Math. Phys. {\bf 72}, no.2,  153--163 (2005)

\bibitem{Araki}
Araki, H.: {\it Relative Hamiltonian for faithful normal states of a von Neumann algebra.} Publ. Res. Inst. Math. Sci. {\bf 9}, 165--209 (1973/74)


\bibitem{AW}
Araki, H., Woods, E.: {\it Representations of the canonical commutation
  relations describing a non-relativistic infinite free bose gas.}
J. Math. Phys. {\bf 4}, 637-662 (1963)


\bibitem{AWyss}
Araki, H., Wyss, W.: {\it Representations of canonical anticommutation relations.} Helv. Phys. Acta {\bf 37}, 136--159  (1964)


\bibitem{BD}
Bruneau, L., Derezi\'nski, J.: {\it Bogoliubov Hamiltonians and one-parameter groups of Bogoliubov transformations}, preprint mp-arc 05-398


\bibitem{BW}
Bisognano, J.J., Wichmann, E.H.: {\it On the duality condition for a Hermitian scalar field.} J. Math. Phys. {\bf 16}, no. 4, 985--1007 (1975)


\bibitem{BFS1}
Bach, V., Fr\"ohlich, J., Sigal, I.M.: {\it Quantum electrodynamics of confined nonrelativistic particles.}  Adv. Math.  {\bf 137}, no. 2, 299--395 (1998)


\bibitem{BFS}
Bach, V., Fr\"ohlich, J., Sigal, I.M.: {\it Return to equilibrium.}  J. Math. Phys.  {\bf 41}, no. 6, 3985-4060  (2000)


\bibitem{BR}
Bratteli, O., Robinson, D.W., Operator Algebras and Quantum Statistical Mechanics I, II. Texts and Monographs in Physics, Springer-Verlag, 1987


\bibitem{DM} DeBi\`evre, S., Merkli, M., in preparation.



\bibitem{DF}
Derezi\'nski, J., Fr\"uboes, R.: {\it  Level shift operator and second order perturbation theory.}  J. Math. Phys.  {\bf 46},  no. 3, 033512, 18 pp.  (2005)


\bibitem{DF1}
Derezi\'nski, J., Fr\"uboes, R.: {\it  Fermi Golden Rule and open quantum systems.}  mp\_arc 05-142, 2005





\bibitem{DJ}
Derezi\'nski, J., Jak\u{s}i\'c, V.: {\it Return to Equilibrium for Pauli-Fierz
  Systems.}  Ann. Henri Poincar\'e  {\bf 4}, no. 4,  739--793 (2003)


\bibitem{DJlso} Derezi\'nski, J., Jak\u{s}i\'c, V.: {\it On the nature of Fermi golden rule for open quantum systems.}  J. Statist. Phys.  {\bf 116},  no. 1-4, 411--423  (2004)


\bibitem{DJP}
Derezi\'nski, J., Jak\u si\'c, V., Pillet, C.-A.: {\it Perturbation theory for $W^*$-dynamics, Liouvilleans and KMS-states.}  Rev. Math. Phys.  {\bf 15}, no. 5,   447--489 (2003)


\bibitem{FM1}
Fr\"ohlich, J., Merkli, M.: {\it Thermal Ionization.}  Math. Phys. Anal. Geom.  {\bf 7},  no. 3, 239--287  (2004)


\bibitem{FM2}
Fr\"ohlich, J., Merkli, M.: {\it Another return of ``Return to Equilibrium''. }  Comm. Math. Phys.  {\bf 251},  no. 2, 235--262  (2004)


\bibitem{FMS}
Fr\"ohlich, J., Merkli, M., Sigal, I.M.: {\it Ionization of atoms in a thermal field.} J. Statist. Phys.  {\bf 116},  no. 1-4, 311--359   (2004)


\bibitem{JP1}
Jak\u{s}i\'c, V., Pillet, C.-A.:  {\it On a model for quantum friction. II. Fermi's golden rule and dynamics at positive temperature.}  Comm. Math. Phys.  {\bf 176},  no. 3, 619--644  (1996)


\bibitem{JP2}
Jak\u{s}i\'c, V., Pillet, C.-A.: {\it On a Model for Quantum Friction
  III. Ergodic Properties of the Spin-Boson System.} Commun. Math. Phys. {\bf
  178}, 627-651 (1996)


\bibitem{JP}
Jak\u{s}i\'c, V., Pillet, C.-A.: {\it Mathematical theory of non-equilibrium quantum statistical mechanics.} J. Statist. Phys.  {\bf 108}, no. 5-6, 787--829. (2002)


\bibitem{JP3}
Jak\u{s}i\'c, V., Pillet, C.-A.: {\it Non-equilibrium steady states of finite quantum systems coupled to thermal reservoirs.}  Comm. Math. Phys.  {\bf 226}, no. 1, 131--162.   (2002)



\bibitem{Marder}
Marder, L.: An Introduction to Relativity. Mathematical Topics, Longmans, Green \& Co Ltd, London and Harlow, 1968



\bibitem{MartinRothen}
Martin, Ph. A., Rothen, F.: Many-Body Problems and Quantum Field Theory. Springer Texts and Monographs in Physics, 2002


\bibitem{M2}
Merkli, M.: {\it Positive Commutators in Non-Equilibrium Quantum Statistical
  Mechanics.} Commun. Math. Phys. {\bf 223}, 327-362 (2001)


\bibitem{M1}
Merkli, M.: {\it The Ideal Quantum Gas. } Lecture notes for the summerschool on open quantum systems, Institut Fourier, Grenoble, 2003


\bibitem{MMS}
Merkli, M., M\"uck, M., Sigal, I.M.: {\it Instability of Equilibrium States for Coupled Heat Reservoirs at Different Temperatures}, submitted


\bibitem{Matthias}
M\"uck, M.: {\it Thermal Relaxation for Particle Systems in Interaction with Several Bosonic Heat Reservoirs}, Books on Demand GmbH, Norderstedt, ISBN 3-8334-1866-4

\bibitem{RSII}
Reed, M., Simon, B.: Methods of modern mathematical physics II, Fourier Analysis, Self-Adjointness. Academic Press 1975


\bibitem{S}
Sewell, G.L.: {\it Quantum Fields on Manifolds: PCT and Gravitationally Induced Thermal States.} Annals of Physics {\bf 141}, 201--224 (1982)

\bibitem{UW}
Unruh, W.G., Wald, R.M.: {\it What happens when an accelerating observer detects a Rindler particle.} Phys. Rev. D, {\bf 29}, No. 6, 1047--1056 (1984)




\end{thebibliography}
\end{document}